\documentstyle[oldlfont,amssymb,eqsecnum,aps]{revtex}

\newtheorem{thm}{Theorem}
\newtheorem{lem}{Lemma} 
\newtheorem{rem}{Remark} 

\newtheorem{problem}{Problem}
\newtheorem{proposition}{Proposition}
\makeatletter

\def\dddot#1{{\mathop{#1}\limits^{\vbox to-1.4pt{\kern-\tw@ pt
 \hbox{\normalshape...}\vss}}}}
\def\ddddot#1{{\mathop{#1}\limits^{\vbox to-1.4pt{\kern-\tw@ pt
 \hbox{\normalshape....}\vss}}}}
\makeatother

\newcommand{\res}{\mathop{\rm res}\nolimits}

\newcommand{\gl}{\mathop{\rm gl}\nolimits}

\newcommand{\Pol}{\mathop{\rm Pol}\nolimits}
\newcommand{\sign}{\mathop{\rm sign}\nolimits}
\newcommand{\ord}{\mathop{\rm ord}\nolimits}

\newenvironment{pf}{ \par\topsep6pt plus6pt
  \trivlist \item[\hskip\labelsep\it 
Proof.]\ignorespaces}{$\Box$\endtrivlist}

\begin{document}
\title{$P_\infty $ algebra of KP, free fermions and
2-cocycle in the Lie algebra of pseudodifferential operators}
\author{A. Yu. Orlov\thanks{Permanent address: Institute of
Oceanology, Krasikova 23, Moscow 117218, Russia}}
\address{Centre de Recherches Math\'ematiques, Universit\'e
de Montr\'eal, Case postale~6128 -- succursale centre-ville,
 Montr\'eal (Qu\'ebec) H3C 3J7, Canada}
\author{P. Winternitz}
\address{Centre de Recherches Math\'ematiques, Universit\'e
de Montr\'eal, Case postale~6128 -- succursale
centre-ville, Montr\'eal (Qu\'ebec) H3C 3J7, Canada}
\maketitle
\begin{abstract}
The symmetry algebra $P_\infty = W_\infty \oplus H \oplus
I_\infty $ of integrable systems is defined. As an example
the
classical Sophus Lie point symmetries of all higher $KP$
equations are obtained. It is shown
 that one (``positive'') half of the point symmetries
belongs to the $W_\infty $ symmetries while the other
(``negative'') part belongs to the $ I_\infty $ ones. The
corresponing action on the tau-function is obtained for
the positive part of the symmetries. The negative part can
not be obtained from the free fermion algebra. A new
embedding of the Virasoro algebra into $gl(\infty )$
describes conformal transformations of the KP time
variables. A free fermion algebra cocycle is described as
a PDO Lie algebra cocycle.
\end{abstract}


\section{Introduction}

In this paper we obtain and investigate the symmetries of
integrable hierarchies based on the use of the algebra of
pseudodifferential operators (PDO). It is a certain
generalization of the famous Kadomtsev-Petviashvili (KP)
hierarchy of partial differential equations (PDE).  We
shall use the word ``symmetry'' in a very general sense: a
``symmetry'' of an equation, or a set of equations, will
be any further equation, compatible with the studied one.
In the context of integrable evolution equations we shall
understand ``symmetries'' as flows, commuting with the
considered flow.

The KP equation~\cite{ref1} was introduced in the context
of waves propagating in shallow water, or in a plasma, and
corresponds to small transverse perturbations of solutions
of the Korteweg-deVries equation. Later it became clear
that applications of the KP equation go far beyond its
first ones. It is currently under intense study in the
context of conformal quantum field theory and string
theory, a partial review of this topic can be found in
Ref.~\cite{ref2}.

The KP equation is known to be integrable, in the sense
that there is a Lax pair associated with it~\cite{ref3},
it allows infinitely many conservation laws, multisolitons
solutions and has all the usual attributes of
integrability~\cite{ref4}--\cite{ref6}.  The KP equation
is the first nontrivial member of an infinite hierarchy of
mutually compatible equations~\cite{ref7,ref8}, each
representing a flow with respect to a different ``time''
$t_n$.

We shall use papers~\cite{ref7} and papers
\cite{ref9}--\cite{ref12}. The famous
$\widehat{\gl}(\infty)$ group symmetry transformations
acting on the $tau-function$ of the KP hierarchy were
presented in Ref.~\cite{ref7}. Here we make the most use
of paper~\cite{ref12}, where symmetries which we shall
refer as ``PDO (pseudodifferential operators) symmetries''
were found. We shall use the notation $P_{\infty}$ for
them in the present paper. The PDO symmetry algebra is a
direct sum of three subspaces: $P_\infty = W_\infty \oplus
H \oplus I_\infty ~$. The subalgebra $H$ is the well-known
Abelian algebra of higher $KP$ flows. The ``positive
part'' of $P_\infty $, now known now as $W_\infty $
algebra symmetries  was exhaustively studied and its
embedding into the $\widehat{\gl}(\infty )$ symmetries was
explicitely described in~\cite{ref9}. The $W_\infty \oplus
H$ subalgebra of KP symmetries results from the
$\widehat{W}_{1+\infty }$ algebra action on the $KP$
tau-function. The $W_{1+\infty } $ symmetries have
numerous applications in matrix models
\cite{ref13}--\cite{ref15},~\cite{ref2},~\cite{ref16}.
For new applications see Ref.~\cite{ref17}. For the
representation theory of the  $\widehat {W} _{1+\infty }$
algebra without connections with the soliton theory
see~\cite{ref18} (for applications in soliton theory only
special weights are available).

The ``negative part'' of $P_\infty $, which we call
$I_\infty $ here, has so far not been adequately studied
in~\cite{ref10} ,~\cite{ref12}.

The other infinite dimensional Lie group associated with
the KP equation is the Lie group of local ``point
transformations'' taking solutions of the equation into
solutions. The Lie algebra of point symmetries forms a
subalgebra in the algebra of all symmetries;  historically
it is this group that is usually called the ``symmetry
group'' of an equation. Point symmetries form classical
object in the theory of partial differential equations,
arising from the work of Sophus Lie, see~\cite{ref19} for
review. The calculation of the ``symmetry group'' of a
differential equation is entirely algorithmic~\cite{ref19}
and can be done using various computer packages
\cite{ref20}.
It turns out that for a number of equations the
corresponding symmetry algebra has a certain typical
structure: it is a semidirect sum of the Virasoro algebra
and a certain nilpotent part of a Kac-Moody one (both are
without central charge)~\cite{ref21,ref22}. This result
was obtained for the KP and Davey-Stewartson equations,
for the 2D Toda lattice, for 3D 3-wave resonant interaction
system and for some other equations
\cite{ref23}-\cite{ref29}. The Virasoro algebra corresponds
to the reparametrization of the $time$ variable. We remark
that it is not the Virasoro algebra corresponding to the
reparametrization of the $spectral~ parameter$ in the
soliton theory~\cite{ref11},\cite{ref12},\cite{ref9},
\cite{ref30,ref31} which was applied to the matrix
models~\cite{ref13}-\cite{ref15} and which completely
belongs to the $W_\infty $ part of $P_\infty $. We shall
show that point symmetries belong to the whole $P_\infty $
and not only to its $W_\infty $ part.

This paper has the following goals: (1) To introduce
the $P_\infty $ algebra. (2) To get all point symmetries
 of the KP and the higher KP equations from the $P_\infty $
 ones and to get the conformal algebra of
 reparametrizations of higher KP time variables.
 (3) To compare the free fermion algebra~\cite{ref7}
 and PDO Lie algebra cocycles~\cite{ref32}. (4) To get
a nonstandard Virasoro algebra from free fermions and
to show that only the positive part of point symmetries
results from free fermion algebra. (5) To calculate the
difference between free fermion Virasoro algebra flows and
 point algebra flows
and to show that this difference corresponds to ``Liouville
 equation flows''~\cite{ref33}.    (6) To consider the
 compatibility of flows and constraints and to explain why
 only a finite part of the infinite-dimensional algebra of
 conformal symmetries survives when one reduces ($3D$)
 higher KP equations to any two-dimensional $KdV-type$
 equation.

We note two facts about point symmmetries: 1. Generally
 they do not preserve
solutions described by the Segal-Wilson Grassmannian~
\cite{ref34} . 2. They create rational solutions which do
 not vanish at infinity~\cite{ref35}.

We mention the following. The KP hierarchy and higher KP
 equations have different representations. The notion of
 ``point symmetry'' depends on the choice of the
 representation. Here we choose the KP higher equation in
its most traditional form as evolutionary
 equation written on one function $w(x,y,t_{N})$ in two
 space variables $x$ and $y$ and one time variable $t_{N}$,
 where N is the number of the equation in the hierarchy.

In Section II we review some results on the KP hierarchy
 and its symmetries, making use of the Gel'fand-Dickey
 approach via the algebra of pseudodifferential operators
 $\cal A $~\cite{ref8} and a space of formal
 Zakharov-Shabat dressings $\cal K $.  We present the
 results in a unified manner and include some known, but
 not easily accessible results~\cite{ref9,ref10,ref11}. We
 introduce a space of generalized KP flows $\cal V $ and
 $P_\infty $ symmetries, $ P_\infty \subset \cal {V} $.

Section III is devoted to the Lie point symmetries of the
 first two equations in the KP hierarchy.

 The relation between the $P_{\infty}$
(and it's subalgebra $W_{\infty}$) and Lie point symmetries
 are derived in Section IV. Two known 2-cocycles are
 compared: the free fermion $\widehat {\gl} (\infty )$
 one~\cite{ref8} and the two-dimensional PDO Lie algebra
 ones~\cite{ref32}.  We pose the compatibility problem for
 pairs of $P_\infty $ constraints and for a constraint and
 flows, and prove that only a finite set of point
 symmetries passes through the $KdV$-type reduction.
 Examples of conditional symmetries are given.

For group times we shall use the following notations: $z ,
 \{t_k \} , \{ t_{mn} \} $.

Except the results in Section IV B this paper is mainly based 
on Ref.~\cite{ref44}.

\section{The generalized $KP$ hierarchy and the 
$P_{\infty}$ symmetries}

\subsection{The space of formal Zakharov-Shabat dressings 
$\cal K$}

We shall make use of the associative algebra of
 pseudodifferential operators (PDO) in one variable on the
 line $a(x, \partial)$ satisfying the permutation rule
\begin{equation}
a(\partial)  b(x) =
	\sum_{k=0}^\infty \frac{b^{(k)} (x) a^{(k)}
 (\partial)}{k!}~ , \label{eq2.1}
\end{equation}
where $(k)$ denotes the $k$-th derivative with respect to
 the argument. For a detailed exposition see e.g.~
\cite{ref8}. There is a natural Lie algebra structure on
 the PDOs given by the commutator $[A,B]=AB-BA$ , we denote 
this algebra by $\cal A $.

An operation of conjugation $(*)$ is introduced, defined by
 the rules
\begin{equation}\label{eq2.2}
x^* = x~,
\quad \partial ^* = - \partial ~,
 \quad (AB)^* = B^* A^*~.
\end{equation}

We shall also make use of the splitting of the space of PDO
 into the direct sum of two linear spaces, $\cal {A} = \cal
{A}_+ \oplus \cal {A}_-$, with

\begin{equation}\label{eq2.3}
( \Sigma a_n \partial^n)_+
	= \sum_{n \ge 0} a_n \partial^n \quad ,\qquad
( \Sigma a_n \partial^n)_-
	= \sum_{n < 0} a_n \partial^n~ .
\end{equation}
We see that $\cal {A}_+$ is the subalgebra of differential
 operators, $A_-$ the subalgebra of purely integral
 operators.

The highest power of the operator $\partial $ in a given
 PDO is called the order of this PDO.

Let us introduce a $space~ of~ formal~ Zakharov-Shabat~
 dressing~ operators$:
\begin{equation}
{\cal K}~=~
 \{ K =
	1 + \sum_{j=1}^\infty K_j (x) \partial^{-j} \}~ .
 \label{eq2.4}
\end{equation}
We call $K$ the formal Zakharov-Shabat dressing operator,
 by analogy with the analytical dressing operator of~
\cite{ref4}.
Points of the space of the formal Zakharov-Shabat dressings
 are parametrized by a  semiinfinite set of arbitrary
 functions of one variable $x$, and sometimes we shall
 write:
\begin{equation}
{\cal K}~=~\{ K_j(x)~,~j=1,2,3,...\infty  \}~.
\end{equation}
Remark. Under certain restrictions one can identify the
 space ${\cal K}$ with the  ``rotated Segal-Wilson
 Grassmannian'' $ g^{-1}W $~\cite{ref34}. We do not
 consider the Grassmannian approach in the present paper.

We consider only sufficiently smooth functions $K_j(x)$,
 then an inverse integral operator exists, namely
\begin{equation}
K^{-1}~=~1 + \sum_{j=1}^\infty \widetilde{K}_j(x)
 \partial^{-j}~, \label{eq2.5}
\end{equation}
with coefficients
\begin{equation}
\widetilde{K}_j(x) =
	-K_j(x) + P(K_1, K_2, \dots , K_{j-1})~,
 \label{eq2.6}
\end{equation}
where $P$ is a differential polynomial in the indicated
 arguments.

We shall also need the $*$-conjugate operator
\begin{equation}
K^{*-1}
	= 1 + \sum_{j=1}^\infty K^*_j (x) \partial^{-j}
 \label{eq2.7}
\end{equation}
(where $K^*_j(x)$ is simply a notation for new
 coefficients, not $*$-conjugates of $K_j(x)$).

In the next subsection we consider certain vector fields
 on the space ${\cal K}$.

\subsection{A space of vector fields $\cal V$. Generalized
 KP flows}

Let us now consider the linear space of pseudodifferential
 operators of the form
\begin{equation}
A_i =
	\sum_n a_n^i(x, z_1, z_2, z_3, \dots)\partial^n \in
 \cal A , \label{eq2.8}
\end{equation}
where $a_n$ are some fixed functions of their arguments.
 The ``time'' $z_i$ plays a privileged role in the operator
 $A_i$ and in the coefficients $a_n^i$.  Note that the
 summation can be over both positive and negative values of
 $n$.  Let us now introduce a mapping
\begin{equation}
A_i \to V_{A_i} \label{eq2.9}
\end{equation}
from the PDOs $A_i$ onto vector fields $V_{A_i}$  on the
 space ${\cal K}$ according to the rule
\begin{equation}
V_{A_i} K \equiv
	\biggl[ \frac{\partial}{\partial z_i}, K \biggr] =
 -(KA_i K^{-1})_- K~. \label{eq2.10}
\end{equation}
We shall denote the space of the vector fields~
(\ref{eq2.10}) by $\cal V $.
One can check that these vector fields keep the chosen form
 of K given by eq.~(\ref{eq2.4}). Each vector field
 $V_{A_i}$ induces a flow of all of the coefficients
$K_n(x)$ of $K$, once we allow the coefficients
 $K_n(x,z_i)$ in eq.~(\ref{eq2.4}) to depend on $ z_i $, as
 well as on $x$. We call these flows the $generalized~ KP~
 flows$ and denote the space of the vector fields~
(\ref{eq2.10}) by $\cal V $.

A result that will be used a great deal below is the
 following.

\begin{thm} \label{thm1} The commutator of $~V_{A_i},
V_{A_j}\in \cal V $ satisfies
\begin{equation}
[V_{A_i}, V_{A_j}]K =
-(KF_{ij} K^{-1})_- K~, \label{eq2.11}
\end{equation}
where we have
\begin{equation}
F_{ij} =
[A_i, A_j] + \frac{\partial A_i}{\partial z_j} -
\frac{\partial A_j}{\partial z_i}~. \label{eq2.12}
\end{equation}
\end{thm}

The proof is a direct calculation, see Ref.~\cite{ref44}.

\newtheorem{sviaz}{Lemma}
\begin{sviaz}
Let $A_i $ be a PDO of the order $p$. The constraint

\begin{equation} \label{restriction}
(KA_iK^{-1})_-=0 \qquad , \qquad K \in \cal K
\end{equation}
restricts  the space ${\cal K}$ to a subspace parametrized
 by $p-1$ arbitrary functions $\{ K_j(x) ~ , ~
 j=1,2,...,p-1 \}$ for $p>0$~. If $p<0$
~(\ref{restriction}) has no solutions.
\end{sviaz}

Remark. One can treat~(\ref{restriction}) as the condition
 of invariance of a  certain subspace of  ${\cal K} $ with
 respect to the flow $V_i$. This condition allows us to
 express $ \{ K_j(x)~,~j\ge p \} $ via $
 \{ K_j(x)~,~j=1,2,...,p-1 \} $
which remains a set of arbitrary functions of one variable.

\newtheorem{guess}{Corollary}
\begin{guess}
Two flows with respect to $z_i$ and to $z_j$ commute iff

\begin{equation} \label{conditional}
(KF_{ij}K^{-1})_-=0~.
\end{equation}

If $F_{ij}(x)$ does not vanish identically this condition
 restricts the space $ {\cal K} $ to a subspace
 parametrized by $p-1$ arbitrary functions $
\{ K_j(x)~,~j=1,2,...,p-1 \} $ for $p>0$, where $p$ is the
 order of $F_{ij}$. The flows do not commute if $p<0$.
\end{guess}

When~ $F_{ij}(x)$ does not vanish identically, but~
(\ref{conditional}) is valid, then we are in the situation
known as $conditional~ symmetries$~\cite{ref29}, namely,
when two flows commute only under some restriction on the
 space of solutions.

\begin{rem} \label{rem1}\normalshape

\begin{enumerate}
\item[1.] The operators $A_i = A^i_{nm} (\vec z)
 x^n\partial^m$ can be viewed as connections corresponding
 to the algebra of PDO and eq.~(\ref{eq2.12}) defines a
 curvature.  Eq.~(\ref{eq2.12}) shows that the mapping
 (\ref{eq2.9}) is not a mapping of the Lie algebra of PDO 
 $\cal A$ 
 to the Lie algebra of vector fields $\cal V$.  The zero
 curvature
 condition $F_{ij} = 0$ guarantees that the flows with
 respect to $z_i$ and $z_j$ are compatible.

\item[2.] A supersymmetric version of eq.~(\ref{eq2.11})
 has been used to construct supersymmetries for the $super~
 KP$ hierarchy~\cite{ref9}. A discrete version of eq.~
(\ref{eq2.11}) was also used in this paper.
\end{enumerate}
\end{rem}

\subsection{The $KP$ hierarchy}

Let us choose $A_n=\partial^n$, where $n$ is any nonzero
 integer. Let us again consider the integral operator $K$
of eq.~(\ref{eq2.4}) and this time interpret the
 coefficients $K_j$ as depending on an infinite sequence of
 times $t_n$.  We shall identify the first
 two as space variables $t_1 = x$, $t_2 = y$, the third
 will be $t_3 = t$.  We keep the notation $\partial \equiv
 \partial_x \equiv \partial_{t_1}$.

We can  write an infinite hierarchy of partial differential
 equations, the Kadomtsev-Petviashvili hierarchy, in the
 following compact form
\begin{equation}
\frac{\partial K}{\partial t_n} =
	-( K\partial^n K^{-1})_- K~. \label{eq2.13}
\end{equation}

It follows from Theorem~\ref{thm1} that the flows
 (\ref{eq2.13}) all commute, i.e.
$[\partial_n, \partial_m]K = 0~, \quad \partial_n = \partial
 \big/ \partial_{t_n}~$. For $n=0$ 
we shall use the different notation $t_{00}$.
As a rule we shall omit the dependence
on the nonpositive times and shall put 
$t_{00}=0,~t_n=0,~ n<0$.
We shall use the notation $\vec t$ for the collection of
 ``KP higher times'' $t_1,t_2,t_3,
\dots $ .
 We put standard
 notations $t_1=x~,~t_2=y$.

Each operator equation (\ref{eq2.13}) (for a fixed $n>0$)
 gives rise to an infinite coupled set of PDEs for the
 coefficients $K_j$ of eq.~(\ref{eq2.4}).

Each equation can be labeled by $(n;i)$:
\begin{equation}
\partial_n K_i - \sum_{\ell = 1}^n  \left({n
\atop \ell-1}\right) K_\ell^{(n - \ell +1)} = \Pol(K_1,
 K_2 , \dots , K_{n+i-2})~, \quad i = 1,2,3,\dots~ ,
 \label{eq2.15}
\end{equation}
where $\Pol$ is a differential polynomial in $K_j~,~
j<n+i-1$ and their $x$-derivatives upto order $n-1$.  The
 linear part of the polynomials has been separated out in
 the left hand side of eq.~(\ref{eq2.15}), so all terms on
 the right hand side are quadratic or higher.  Note that
 each equation in the system (\ref{eq2.15}) involves just
 one time $t_n$ and the variable $x$.

$ How~ does~one~ get~ the~ three~ dimensional~ KP~and~
 higher~ KP~ equations~from~(\ref{eq2.13})?~ $ These
 equations are written for a set $\{ K_i~,~1 \le i \le r
 \}$
 as unknown functions of $d$ variables $t_{n_k}~,~
k=1,2,...,d$, which play the role of space (time) variables
 in $d$-dimensional space. They are obtained by taking the
 set of equations (\ref{eq2.15}) with labels $(n_k;i)$.
To get equations in the three dimensional space spanned by
 $~x,t_m,t_n~$ one should consider a system of $r$
 equations for $\{ K_i~,~1 \le i \le r \}~,~r=m+n-2$. In
 the present paper we shall consider KP higher equations in
 3-dimensional space spanned by $~x,y,t_N~$ and we shall
 call this equation $KP_N$.

For example, to obtain the Kadomtsev-Petviashvili equation
 itself, or $KP_3$, take $n=3$~and $n=2$ and hence $(n;i) =
 (3;1)$, $(2;1)$ and $(2;2)$.  The corresponding equations
 are
\begin{equation}
\partial_3 K_1 - (K_1^{'''} + 3K_2^{''} + 3K_3')
	= 3K_1^2 K_1' - 3K_1K_1^{''} - 3K_1^{\prime 2} -
 3K_2' K_1 - 3K_1'K_2~ , \label{eq2.17}
\end{equation}
\begin{equation}
\partial_2K_1 - (K_1^{''} + 2K_2')
	= -2K_1K_1'~ ,\quad
\partial_2K_2 - (K_2^{''} + 2K_3')
	= -2K_1'K_2~. \label{eq2.18}
\end{equation}
All three are evolution equations (in $t_3$, $t_2$ and
 $t_2$, respectively) for the functions $K_1$, $K_2$ and
 $K_3$.  Using (\ref{eq2.18})
to eliminate $K_3$ and $K_2$ from eq.~(\ref{eq2.17}), we
 obtain the KP equation
\begin{equation}
w_{xt_3}
	= \frac{1}{4} w_{xxxx} + \frac{3}{2} w_x w_{xx} +
 \frac{3}{4} w_{yy}~,\qquad
w(\vec t)
	= -2K_1(\vec t) \label{eq2.20}
\end{equation}
(the usual KP equation is satisfied by $u \equiv w_x$).

Next example is $KP_4$. It is obtained by taking $n=4$
 and hence $(n;i) = (4;1)$, $(3;1)$, $(3;2)$, $(2;1)$,
$(2;2)$,
$(2;3)$. One gets:
\begin{equation}
w_{xxxxy} + 6w_{xx} w_{xy} + 4w_x w_{xxy} + 2w_{xxx} w_y -
 2w_{xxt_4} + w_{yyy} = 0~. \label{eq2.26}
\end{equation}

In what follows we shall use the
 evolutionary equation 
which describes KP flow~(\ref{eq2.13}) on the
function $w = -2K_1(x, y , t_3,\dots)$
\begin{equation}
\frac{\partial w}{\partial t_n} = 2 \res_\partial
 K\partial^n K^{-1} 
 \label{eq2.27}
\end{equation}
as the $KP_n$ equation.

\subsection{Formal Baker-Akhiezer functions}

We shall make use of the formal Baker-Akhiezer functions~
\cite{ref8}.  They can be defined in terms of the PDO $K$
 of eq.~(\ref{eq2.4}) by putting
\begin{mathletters}\label{eq2.30}
\begin{equation}
\varphi(\lambda, t_0,\vec t)
	= Ke^\zeta = e^\zeta \biggl( 1 + \sum_{n=1}
 \lambda^{-n} K_n (t_0,\vec t) \biggr)~, \label{2.30a}
\end{equation}
\begin{equation}
\varphi^*(\lambda,t_0, \vec t)
	= K^{*-1} e^{-\zeta} = e^{-\zeta} \biggl( 1 +
 \sum_{n=1} \lambda^{-n} K_N^*(t_0,\vec t) \biggr)~,
 \label{2.30b}
\end{equation}
\begin{equation}
\zeta(\lambda,t_0,\vec t) = \sum_{k=1} \lambda^k t_k +
t_0\ln\lambda \quad , \quad \vec t = t_1, t_2, \dots ~,
 \quad t_1 = x~, \label{2.30c}
\end{equation}
\end{mathletters}\ignorespaces
where $\lambda$ is a formal complex parameter. We introduce
 here the additional  parameter $t_0$ as 
in~\cite{ref30},~
\cite{ref31} ($t_0$ is not $t_{00}$!). It is responsible for
 the Schlesinger
discrete transformations - special cases of Darboux
 transformations of the $KP$ hierarchy. It was not used in~
\cite{ref8}.

We need the following lemma, due to Dickey~\cite{ref8}.

\begin{lem} Let $P = \sum_k p^k \partial^k$ and $Q = \sum_k
 q^k \partial^{-k}$ be any two PDOs and let
\begin{equation}
\res_{\partial} \Sigma a_n \partial^{-n} = a_{1}~.
 \label{eq2.31}
\end{equation}
Then we have
\begin{equation}
\res_{\lambda} (Pe^{\lambda x}) (Qe^{-\lambda x}) =
 \res_\partial PQ^*. \label{eq2.32}
\end{equation}
\end{lem}

Using the above lemma, eq.~(\ref{eq2.27}) and~
(\ref{eq2.30}) we obtain a formula summing up the KP
 hierarchy in terms of the formal Baker-Akhiezer functions:
\begin{equation}
\frac{\partial w(t_0,\vec t)}{\partial t_m}
	= 2 \res_\lambda \lambda^m \varphi (\lambda ,
t_0,\vec t) \varphi^*(\lambda ,t_0,\vec t)~, \label{eq2.33}
\end{equation}
which can be rewritten as
\begin{equation}
2 \varphi \varphi^* =
	2 + \frac{w_x}{\lambda^2} + \frac{w_y}{\lambda^3} +
 \frac{w_t}{\lambda^4} + \dots~. \label{eq2.34}
\end{equation}
The asterisk in eq.~(\ref{eq2.34}) has the same meaning as
 in eq.~(\ref{eq2.7}), i.e. it does not indicate
 conjugation.
The term $\lambda^{-1}$ is absent in eq.~(\ref{eq2.34}),
 since we have $K_1 + K_1^* = 0$.  The $\lambda^0$ term on
the right hand side of eq.~(\ref{eq2.34}) follows from the
 asymptotic behavior of the Baker functions for $\lambda
\to + \infty$.

In addition to~(\ref{eq2.33}) we consider 'Schlesinger
 transformation' as a discrete equation with respect to
 $t_0$~\cite{ref36},\cite{ref30}:

\begin{equation}\label{Schloesinger}
w(t_0+1,\vec t )-w(t_0,\vec t )~=~2\partial \ln \varphi
 (0,t_0,\vec t ) ~.
\end{equation}

\subsection{ $P_{\infty}$ symmetries of the KP hierarchy}

We call a ``symmetry'' of the KP equation any
 differential, or integrodifferential equation that is
 evolutionary in group time $z$ i.e. of the form
\begin{equation}
\frac{\partial w}{\partial z} = F[w, \vec t , z],
 \label{eq2.35}
\end{equation}
where $F$ is a function of $w$, its $x$ and $y$ derivatives
 and of integrals of the type $\partial^{-k} w$, $k>0$,
 that is compatible with the KP equation itself.
 Similarly, a symmetry of the KP hierarchy will be any
 equation of the form (\ref{eq2.35}), compatible with the
 entire hierarchy. That is for each $m$
\begin{equation}\label{opredeleniesymmetry}
\partial _z \partial _{t_m} w = \partial _{t_m} \partial _z
 w~.
\end{equation}

All higher equations in the KP
 hierarchy are Abelian symmetries of
 the KP equation itself.
Other symmetries of the KP equation exist and, contrary to
 the KP hierarchy, typically the corresponding equations
 (\ref{eq2.35}) have coefficients explicitly depending on
 the independent variables $x$, $y$, $t$.  Among them we
 mention the Lie points symmetries~\cite{ref21,ref22}.  The 
most complete
 treatment of all symmetries of the KP hierarchy in the
 framework of integrability theory is given in Ref.~
\cite{ref9},\dots ,~\cite{ref12}.  To put those results
 into the present context, let us use the mapping of
 Section~II.B from the space of pseudodifferential
 operators to vector fields.

First let us consider the following extension:
\begin{equation}\label{xextension}
x\quad \to \quad \hat x =~x~+~ t_0\partial^{-1}+
\sum_{k\neq 1} k t_k \partial^{k-1}~.
\end{equation}
The notation $ \Box $ was used in Ref.~
\cite{ref9},\cite{ref10},\cite{ref12} rather than $\hat x $, 
and there was $t_0 \equiv 0$.
We choose the PDOs to be
\begin{equation}
A_{mn} = \hat x^n \partial^m \quad , \qquad n,m \in \Bbb Z
\label{eq2.36}
\end{equation}
and construct the mapping $A_{mn} \to V_{mn}$ with
\begin{equation}\label{eq2.37}
V_{mn} K = [\partial_{mn}, K] = -(K\hat x^n \partial^m
 K^{-1})_- K~,
\end{equation}
where $\partial _{mn} $ is a derivative with respect to the
 group time $t_{mn}$. We have $t_{m0} \equiv 
t_m~,~m \neq 0$ 
($t_{00}$ is not $t_0$). 
It is necessary to define negative
 powers of $\hat x$ to represent them as a power series in
 $\partial $. This can be done in many different ways. In
 the present paper we shall define them in the two ways. We
 shall consider that $KP$ higher times
vanish starting from a certain number $N+1$, where $N $ is
 the order of $\hat x $. Then
we put
\begin{equation} \label{i}
(\hat x)^{-1}=(Nt_N)^{-1}(\partial)^{1-N}(1+O(\partial
 ^{-1}))~,
\end{equation}
which is a purely integral operator of order $-N$.
For other applications we take $t_k=0,~k<0$ 
and put
\begin{equation} \label{ii}
(\hat x)^{-1}=
(1+O(\partial ))(t_0+1)^{-1}\partial~,
\end{equation}
which is a differential operator of infinite order, the
 value of $t_0 $ is noninteger.

For any given choice of $\hat x ^{-1}$ due to the extension
~(\ref{xextension}) it follows from Theorem~\ref{thm1} that
 we have
\begin{equation}
(\partial_{mn} \partial_k - \partial_k \partial_{mn})K = 0~
, \qquad n,m\in \Bbb Z~, \label{eq2.38}
\end{equation}
and hence for any values of $m$ and $n$ the flow of
 (\ref{eq2.36}) is compatible with the flows of the KP
hierarchy.

The Lie algebra of the vector fields $V_{mn}$, generated by
 $\hat x^n \partial_n^m$ for $n \ne 0$, $n \in \Bbb Z$ and
 $\partial_{t_m} - \partial_x^m$ for 
$n=0,~m>0$, is the algebra
 of ``additional symmetries'' of the KP hierarchy
 introduced in~\cite{ref12},\cite{ref10}. Now we
 consider  $m,n,m',n' \in \Bbb Z $. 
The commutation relations have the form:

\begin{mathletters}\label{eq2.39}
\begin{equation}
[V_{mn}, V_{m'n'}]~
 \sim ~[x^n
 \partial^m, x^{n'} \partial^{m'}]\quad, \quad nn' \ne 0
~, \label{eq2.39a}
\end{equation}
\begin{equation}
\qquad [V_{mn}, V_{m'n'}]~ =~0 \quad ,\quad \qquad \qquad
nn' = 0~. \label{eq2.39b}
\end{equation}
\end{mathletters}\ignorespaces

The algebra of vector fields $V_{mn}$ will be called the
 $P_{\infty}$ algebra.
It is a direct sum of three subspaces $P_\infty = W_\infty
 \oplus H \oplus I_\infty $. The subalgebra of vector
 fields $~V_{mn}~,~n>0~$ is a $W_\infty $ algebra. The
 case$~n=0$ describes the higher $KP$ flows. Then
 ``negative'' subalgebra $n<0$ will be referred to as
 $I_\infty $.
Note that due to the last commutation relation~
(\ref{eq2.39}) this algebra is different from the Lie
 algebra of PDOs in one variable $\cal A$  described in 
Section II A.

Until a certain convention on what is $\hat x ^{-n}$ is
 adopted, the symmetry action of the $I_\infty $ algebra on
 the KP solutions is undefined (it was undefined in~
\cite{ref12}). We also note that to get $P_\infty $
symmetries from the free fermion algebra
$\widehat{gl}(\infty)$ action on tau-functions, we choose
 the convention~(\ref{i}) and keep the parameter $ t_0 $ as
 noninteger.

$P_\infty $ has infinitely many different infinite
 dimensional Abelian subalgebras. Any one of them can be
 chosen to generate a hierarchy of commuting flows in
 addition to higher KP flows (see Remark 7 in~\cite{ref9}). 
They could
be called {\em ``second-level KP hierarchies''}.
 Each hierarchy $KP[Q]$ is defined by a PDO $Q(\hat x
 ,\partial)$,
which produces the flows:
\begin{equation}\label{secondlevel}
V_{Q^n} \in P_\infty \quad , \quad n \in \Bbb Z
\end{equation}
with respect to the times $t^{(Q)}_n$~. For $Q=\partial $
 we get the KP hierarchy itself. Both hierarchies commute
 due to the extension $x \to \hat x$. As in the case of the
 KP hierarchy, it is possible to construct finite closed
 subsystems of $r$ partial differential equations for a
 subset of the functions $\{ K_i(x)~,~ i=1,2,...,r<\infty
 \} $ if $\ord~Q<\infty $.  As in Section~II.C it is then
 possible to construct evolution equations for one function
 $w$, evolving in a 3-dimensional space, spanned by ($x$,
 $t^{(Q)}_m$, $t^{(Q)}_n$) with any integers $m,n$.

Also the algebra $P_{\infty }$ of flows contains infinitely
 many different sets of Virasoro flows:

\begin{equation} \label{virflows}
V_{Q^{n+1}P} \in P_\infty \quad ,\quad Q=Q(\hat x ,\partial
 ) \quad ,\quad P=P(\hat x ,\partial ) \quad ,\quad [P,Q]~=
~1~.
\end{equation}
\newtheorem{highlevel}{Remark}.
\begin{highlevel}
 In the same way as the $P_\infty $ algebra for the KP
 hierarchy, one constructs the $P_\infty $ algebra for the
 $KP[Q]$~hierarchy if an invertible $\hat P $ exists. One
 considers flows $V_{\hat P ^nQ^m}$ similar to ~
(\ref{eq2.36})-(\ref{eq2.37}) with the change
 $(\hat x,\partial ) \to (\hat P ,Q)$, where $\hat P$ is
 the extension of $P$:
\begin{equation}\label{hernia}
\
\hat P~=~P~+~t_0^{(Q)}Q^{-1}~+~\sum kt_k^{(Q)}Q^{k-1} \quad
 ,\quad k \in \Bbb Z~.
\end{equation}
These second-level $P_\infty $ flows commute with both KP
 and KP[Q] hierarchies. This way one can produce the n-th
 level KP hierarchy and its $P_\infty $ flows.
\end{highlevel}

We shall not discuss these higher-level $P_\infty $ in the
 present paper.

 Amongst the Virasoro flows we mention one of particular
 interest. It corresponds to $V_{\hat x \partial
 ^{m}}\equiv V_{m1}~\in~W_\infty $. It was introduced in
 Ref.~\cite{ref9}--\cite{ref12}, and corresponds to a
 reparametrization of the spectral parameter $\lambda $.
 It has been used for establishing relations between
 soliton theory and quantum field theory in~
\cite{ref30,ref13}. These flows are completely in the
 $W_\infty $ part of $P_\infty $ symmetries.

Example. We shall consider below the following Virasoro
 flows $V_{P^{n+1}Q} \in W_\infty ,~H,~I_\infty $ for $~
n>-1~,~n=-1~,~n<-1~$ respectevely. As we shall see for
$Q=\partial ^N$ these flows correspond to the
 reparametrizations of a higher KP time variable $t_N$. For
 $N=3,4$ these flows result in the point symmetries of
 $KP_3$ and $KP_4$ equations which first were obtained in~
\cite{ref21,ref22} with the help of a computer program.

~

Following~\cite{ref9} we introduce the pair
\begin{equation} \label{LM}
L = K \partial K^{-1}~~  ,~~   M = K \hat x K^{-1}~~  ,~~
  [L,M]=1~ ,
\end{equation}
which act in the following way on the formal Baker function:
\begin{equation}
 L\varphi (\lambda )=\lambda \varphi (\lambda )~~~,~~
~M\varphi (\lambda) = \frac{\partial
 \varphi(\lambda)}{\partial \lambda}~.
\end{equation}

Let us define $M^{-1}=K\hat x ^{-1}K^{-1}$ and the
 corresponding $\partial_{\lambda}^{-1}$. We write formally:

\begin{equation} \label{neg act}
M^{-1}\varphi (\lambda) = \partial_{\lambda}^{-1}\varphi
(\lambda)~.
\end{equation}

In agreement with eq.~(\ref{ii}) we treat $\partial
_{\lambda}^{-1}$
in the following way: we expand $ e^\zeta $ in the Baker
 function~(\ref{eq2.30}) into a positive power series in
 $\lambda$ and formally multiply it by the rest $O(1)$ part
 of $\varphi$ . Then we integrate
 each term according to the following rule:
\begin{equation}\label{agreementii}
\partial _{\lambda } ^{-1}\lambda ^{n + t_0} = (n + t_0 +
 1)^{-1}\lambda ^{n + t_0 + 1} \quad ,\quad n \in \Bbb Z~,
\end{equation}
where we should take a noninteger value of $t_0$. This
 convention is available to embed $P_{\infty}$ symmetries
 into the
Japanese fermionic $\widehat {gl} (\infty)$ ones, see below.

In agreement with~(\ref{i}) we treat $\partial
_{\lambda}^{-1}$ in eq.~(\ref{neg act}) as a path integral
 over $\lambda $ from the point $\lambda $ to $\lambda ~=~
\infty $:
\begin{equation}\label{agreementi}
\partial_{\lambda}^{-1}\varphi (\lambda ,t_0,\vec t )~=
\int ^\lambda _\infty \varphi (\lambda ',t_0,\vec
t)d\lambda '~,\quad \vert {\vec t} \vert \ne 0~,
\end{equation}
where the path is so chosen that for $\lambda ~\to~ \infty
 $ , the integrand vanishes.
This convention will be used below to obtain point
 symmetries. For each convention~
(\ref{agreementi}),(\ref{agreementii}) $\partial _\lambda
 ^{-n}~=~(\partial _\lambda ^{-1})^n$.

The flow of the function $w = -2K_1$ with respect to the
 ``time'' $t_{mn}$ of eq.~(\ref{eq2.37}) is given by the
 following theorem.

\begin{thm} \label{thm2}
Let $K$ be the PDO of eq.~(\ref{eq2.4}) and $t_{mn}$, where
 $m,n$ are any integers, the time defined in eq.~
{\em(\ref{eq2.36})}, {\em(\ref{eq2.37})}. Given the
 convention about $n<0$, the $P_\infty $ flows of $w =
 -2K_1$ with respect to the times $t_{mn}$ are given as
\begin{mathletters}\label{eq2.40}
\begin{equation}
\partial_{mn} w = 2 \res_\partial (K \hat x^n \partial^m
 K^{-1}) = 2 \res_\partial M^nL^m, \label{eq2.40a}
\end{equation}
or equivalently
\begin{equation}
\partial_{mn} w
	= 2 \res_\lambda \lambda^m \frac{\partial^n
 \varphi(\lambda)}{\partial \lambda^n} \varphi^*(\lambda).
 \label{eq2.40b}
\end{equation}
\end{mathletters}\ignorespaces

\end{thm}

\begin{pf}  Theorem~\ref{thm2} is a consequence of the
 Lemma of Section~II.D.  We set $P = K \hat x \partial^m$,
 $Q = K^{*-1}$ and use the definition (\ref{eq2.30}) of the
 formal Baker-Akhiezer functions.  We then have
\begin{equation}
\res_\partial K \hat x^n \partial^m K^{-1}
	= \res_\lambda(K \hat x^n \partial^m
 e^\zeta)(K^{*-1} e^{-\zeta})
	= \res_\lambda \lambda^m (K \hat x^n
 e^\zeta)(K^{*-1} e^{-\zeta})
	= \res_\lambda \lambda^m \frac{\partial^n
\varphi(\lambda)}{\partial \lambda^n} \varphi^*(\lambda).
\end{equation}
\end{pf}

\subsection{Symmetries via vertex operators}

In order to link $\widehat{\gl}(\infty)$ symmetries~
\cite{ref7} which act on $\tau$-function with ``PDO''
 ones,\cite{ref9}, \dots ,~\cite{ref12} we need some
 results on vertex operators and $\tau$-functions.  The
 vertex operators can be written as
\begin{equation}
X(\lambda ,t_0,\vec t )
	= \exp \biggl(t_0 \ln \lambda + \sum_{k=1}^\infty
 \lambda^k t_k \biggr) \exp \biggl( -\partial_0 - \sum
_{k=1}^\infty \frac{1}{k \lambda^k} \partial_k \biggr)~,
 \label{eq2.41}
\end{equation}
\begin{equation}
X^*(\lambda ,t_0,\vec t ) = \exp(-t_0 \ln \lambda - \sum
_{k=1}^\infty \lambda^k t_k \biggr) \exp \biggl( \partial
_0 + \sum_{k=1}^\infty \frac{1}{k \lambda^k} \partial_k
\biggr)~. \label{eq2.42}
\end{equation}

The zero mode $t_0 \ln \lambda - \partial_0$ was added to
 the vertex operator in Ref.~\cite{ref30} to simplify
 calculations.  It can be checked by a direct calculation
 that the vertex operators introduced above satisfy the
 fermion algebra anticommutation relations:

\begin{mathletters}\label{eq2.43}
\begin{equation}
X(\lambda) X(\mu) + X(\mu) X(\lambda) = 0~, \label{eq2.43a}
\end{equation}
\begin{equation}
X^*(\lambda) X^*(\mu) + X^*(\mu) X^*(\lambda) = 0~,
 \label{eq2.43b}
\end{equation}
\begin{equation}
X(\mu) X^*(\lambda) + X^*(\lambda) X(\mu) = \delta (\lambda
 - \mu)~, \label{eq2.43c}
\end{equation}
\end{mathletters}\ignorespaces
where $\delta(\lambda - \mu)$ is the Dirac
$\delta$-function with respect to integration about a
 circle $S^1$ (close to $\lambda \to \infty$)
\begin{mathletters}\label{eq2.44}
\begin{equation}
\delta(\lambda - \mu) = \sum_{n=-\infty}^\infty \biggl(
 \frac{\mu}{\lambda} \biggr)^n \frac{1}{\lambda}~,
 \label{eq2.44a}
\end{equation}
\begin{equation}
\oint f(\lambda) \delta(\lambda - \mu)d \lambda = f(\mu)~.
 \label{eq2.44b}
\end{equation}
\end{mathletters}\ignorespaces
The ``zero-time'' $t_0$ as a discrete variable, was
 introduced in Ref.~\cite{ref7} and~\cite{ref36}, though it
 was not used in these papers to add a zero mode to the
 vertex operators (\ref{eq2.41}) and (\ref{eq2.42}).  One
 of the uses it was put to in Ref.~\cite{ref30} was to
 introduce the flag space of Grassmannians into KP theory.

A $\tau$-function is a function of all the times $\vec t =
 \{ t_1, t_2, \dots \}$ and also of the discrete ``zero''
 time  $t_0$~\cite{ref7,ref36,ref30,ref31}
\begin{equation}
\tau
	= \tau_n (\vec t)~, \quad n = t_0~. \label{eq2.45}
\end{equation}

The formal Baker function near $\lambda = \infty$ can be
 expressed in terms of vertex operators and the
 $\tau$-function as
\begin{equation}
\varphi(\lambda, t_0, \vec t)
	= \frac{X(\lambda, t_0, \vec t) \tau(t_0 + 1,
\vec t)}{\tau(t_0, \vec t)} \label{eq2.46}
\end{equation}
(with a similar expression for $\varphi ^*$).

For any sufficiently small shift 
$(\vec t - \vec {t'})$ we
 have the bilinear identity
\begin{equation}
\res_{\lambda = \infty} \varphi( \lambda, t_0, \vec t)
 \varphi^*(\lambda, t_0, \vec t') = 0~. \label{eq2.47}
\end{equation}

Let us now consider variations of the $\tau$-function due
 to the vector field
\begin{equation}
V_{\lambda \mu } \tau
= X(\lambda ) X^*(\mu ) \tau~. \label{eq2.48}
\end{equation}

In the Kyoto school approach~\cite{ref7} this is an
 infinitesimal group transformation of the $\tau$-function,
 corresponding to an action of the algebra $\widehat
 {\gl}(\infty)$.  Let us calculate the corresponding
 induced action on KP solutions $w_1$, following ref.~
\cite{ref9}.  We have
\begin{equation}
V_{\lambda \mu} w
	= -2 \res_{k=\infty} \bigl(V_{\lambda \mu}
 \varphi(k) \bigr) \varphi^*(k)~, \label{eq2.49}
\end{equation}
where $k$ is a spectral parameter and all times in
 $\varphi(k)$ and $\varphi^*(k)$ are set equal.  In
 deriving (\ref{eq2.49}) use is made of the relations
\begin{equation}
\varphi(k)
	= e^{\zeta(k)} \biggl( 1 - \frac{w}{2k} + O
 \biggl( \frac{1}{k^2}\biggr)\biggr)\quad,
\quad
\varphi^*(k)
	= e^{-\zeta(k)} \biggl( 1 + O \biggl( \frac{1}{k}
 \biggr)\biggr)~. \label{eq2.50}
\end{equation}
We recall that the relation between the $\tau$-functions
 and solutions is
\begin{equation}
w(t_0,\vec t)
	= 2 \frac{\partial}{\partial x} \ln \tau (t_0,\vec
t)~, \label{eq2.51}
\end{equation}
but we do not use that here.  For the Baker function we
 have
\begin{equation}
V_{\lambda \mu} \varphi(k)
	= \frac{X(k)(V_{\lambda\mu} \tau)}{\tau} -
 \varphi(k) \frac{V_{\lambda \mu} \tau}{\tau}~.
\label{eq2.52}
\end{equation}
The last term in eq.~(\ref{eq2.52}) does not contribute to
 the residue in eq.~(\ref{eq2.49}) and we obtain
\begin{equation}
V_{\lambda\mu}w
	= -2 \res_k \frac{\bigl(X(k)
 X(\lambda)X^*(\mu)\tau\bigr)}{\tau} \varphi^*(k)~.
 \label{eq2.53}
\end{equation}

From the fermion commutation relations (\ref{eq2.43}) we
 obtain
\begin{equation}
\bigl[X(k), X(\lambda)X^*(\mu)\bigr]
	= -\delta(k - \mu) X(\lambda)~, \label{eq2.54}
\end{equation}
and hence
\begin{equation}
- \res_k \frac{\bigl(X(k) X(\lambda)
 X^*(\mu)\tau\bigr)}{\tau} \varphi^*(k)
	= \res_k \delta(k - \mu)
 \frac{X(\lambda)\tau}{\tau} \varphi^*(k) - \res_k
\frac{X(\lambda) X^*(\mu) X(k)\tau}{\tau} \varphi^*(k)~.
 \label{eq2.55}
\end{equation}

Making use of the bilinear identity (\ref{eq2.47}) we can
 see that the last term in eq.~(\ref{eq2.55}) vanishes.

Using eq.~(\ref{eq2.53}), (\ref{eq2.54}) and
 (\ref{eq2.46}), we obtain the final result
\begin{equation}
V_{\lambda\mu} w
	= 2 \varphi(\lambda) \varphi^*(\mu)~.
 \label{eq2.56}
\end{equation}
This formula was announced in Ref.~\cite{ref12} (with some
 obvious misprints) and a proof was given in Ref.~
\cite{ref9}.  Eq.~(\ref{eq2.56}) plays a key role if we
 wish to relate the symmetries (\ref{eq2.40}) with the
$\widehat{\gl}(\infty)$ transforms~\cite{ref7}.  Using eq.~
(\ref{eq2.48}) and (\ref{eq2.56}) we can write the
following commutative diagram


\begin{equation}
\matrix{
V_{\lambda\mu} \tau
	= X(\lambda) X^*(\mu)\tau &\longrightarrow&
 V_{\lambda\mu} w = 2\varphi(\lambda) \varphi^*(\mu)\cr
\downarrow && \downarrow\cr
\partial_{mn}\tau
	=\res \lambda^m \frac{\partial^n
 X(\lambda)}{\partial \lambda^n} X^*(\lambda)\tau
 &\longrightarrow& \partial_{mn}w = 2 \res \lambda^m
 \frac{\partial^n\varphi}{\partial \lambda^n} \varphi^*~.  }
 \label{eq2.58}
\end{equation}

Remark. In~(\ref{eq2.58}) we have  $n \ge 0$. The second
 formula defines the action of $\widehat{W}_{1+\infty}$
 generators acting on the tau-function.
The case $n<0$ is considered in detail in the forthcoming
 paper.

Below, in Section IV, we shall make use of the formalism of
 Sections~II.E and II.F to obtain all point symmetries of
 the equations of the KP-hierarchy.

\section{Point symmetries of the KP equations}

Point symmetry is the following particular case of general
 symmetry~(\ref{eq2.35}):
\begin{equation}\label{point}
\frac {\partial w}{\partial z} = (\vec v \cdot \vec
 \partial )w - p \quad , \quad \partial = (\partial _x ,
 \partial _y , \partial _{t_N} )~,\quad \vec v =
 (v_1,v_2,v_N) \quad ,\quad v_i = v_i(w,x,y,t_N)~,\quad p =
 p(w,x,y,t_N)~.
\end{equation}

This equation looks like one which appears in the
problem of a passive scalar (see the paper of 
 I.V.Kolokolov in the present book).

We shall see that in our case $p,v_i$
 are parametrized by five (or less) arbitrary
functions $f,g,h,k,\ell $ of one variable.

The point symmetry algebra of (\ref{eq2.20}) was calculated
 in Ref.~\cite{ref21}, it is a sum of the vector
fields
\begin{equation}\label{eq3.2}
\hat V
	= T(f) + Y(g) + X(h) + W(k) + U(\ell),
 \label{KP3point1}
\end{equation}
which act in four-dimensional space spanned by $x,y,t,w$.
 In terms of flows we have (we add $t_0$ term here):
\begin{equation}
w_z = fw_t + \biggl( \frac23 yf' + g \biggr) w_y +
 \biggr[ \frac13 xf' + \frac29 y^2f{''} + \frac23 yg' + h
\biggr]w_x + \label{KP3point2}
\end{equation}
$$
\biggl[ \frac13 wf' + \frac19 (x^2+4yt_0)f{''} + \frac4{27}
 xy^2 f{'''} + \frac{4}{243} y^4 f{''''} + \frac23t_0g{'} +
  \frac49 xyg{''} - \frac{8}{81} y^2 g{'''} + \frac23 xh' -
 \frac49 y^2 h{''} - yk - \ell \biggr]~.
$$

The vector fields (\ref{KP3point1}) can be integrated to
 yield point transformations, taking solutions of the PKP
 equation into solutions~\cite{ref21,ref22}.

The nonzero commutation relations of the symmetry algebra
 are
\begin{equation}
\bigl[T(f_1), T(f_2)\bigr]
	= T(f_1f_2' - f_1'f_2)~, \label{eq3.4}
\end{equation}
\begin{mathletters}\label{eq3.5}
\begin{equation}
\bigl[Y(g_1), Y(g_2)\bigr]
	= \frac23 X(g_1g_2' - g_1'g_2)~, \label{eq3.5a}
\end{equation}
\begin{equation}
\bigl[Y(g), X(h)\bigr]
	= \frac49W(hg{''} - g'h') - \frac89 U(gh{''})~,
\label{eq3.5b}
\end{equation}
\begin{equation}
\bigl[Y(g), W(k)\bigr]
	= U(gk)~,  \quad
\bigl[X(h_1), X(h_2)\bigl]
	= -\frac23 U(h_1h_2' - h_1'h_2)~,  \label{eq3.5c}
\end{equation}
\end{mathletters} \ignorespaces
\begin{mathletters}\label{eq3.6}
\begin{equation}\label{eq3.6a}
\bigl[T(f), Y(g)\bigr]
= Y\biggl(f\dot g - \frac23 g \dot f\biggr)~,
\quad \bigl[T(f), X(h)\bigr]
= X\biggl(f \dot h - \frac13 h \dot f\biggr)~,
\end{equation}
\begin{equation}\label{eq3.6b}
\bigl[T(f), W(k)\bigr]
	= W(f \dot k + \dot f k)~,\quad
\bigl[T(f), U(\ell)\bigr]
= U\biggl(f \dot \ell + \frac13 \ell
\dot f\biggr).
\end{equation}
\end{mathletters}\ignorespaces
We see that the fields $T(f)$ form a centerless conformal
 algebra.  The vector fields $\{ Y(g), X(h), W(k), U(\ell)
 \}$ form a certain nilpotent subalgebra of centerless
 Kac-Moody algebra. This is a loop algebra with $t$ as the
 loop parameter.

For $KP_4$ the algebra of point symmetries is again a
 semidirect sum of  a Kac-Moody and Virasoro algebras:
\begin{equation}
V
	= T(f) + Y(g) + X(h) + U(\ell)~, \label{KP4point1}
\end{equation}
with
\begin{mathletters}\label{eq3.12}
\begin{equation}
T(f)
	= f \partial_{t_4} + \frac14 \dot f(x \partial_x +
 2y \partial_y) - \frac14 (w \dot f + xy \ddot f) \partial
_w~, \label{KP4T}
\end{equation}
\begin{equation}\label{KP4X}
Y(g)
	= g\partial_y - \frac12 x \dot g \partial_w~,
 \quad
X(h)~= h\partial_x - y \dot h \partial_w~,
 \quad
U(\ell)~= \ell \partial_w~,
\end{equation}
\end{mathletters}\ignorespaces
where $f(z)$, $g(z)$, $h(z)$ and $\ell(z)$ are all arbitrary
 functions of the time $t_4 = z$.

An earlier observation is that all known integrable PDEs in
 3-dimensions have Kac-Moody-Virasoro algebras as Lie point
 symmetry algebras.  To the KP, equation~
\cite{ref21,ref22}, the Davey-Stewartson equation~
\cite{ref23}, the 3 wave resonant interaction equations~
\cite{ref24} and several others, we have just added the
 higher order KP equation (\ref{eq2.26}).

In Section~IV we shall show that the same is true for each
 equation in the  KP hierarchy.

\section{Lie point symmetries,
 $P_\infty $ ones and $\widehat{\gl}(\infty)$
 algebra}
\subsection{Extraction of point symmetries}

Let us obtain the point symmetries for higher $KP$
 equations and show that the corresponding algebras have
 a Kac-Moody-Virasoro structure.  The corresponding Lie
 point symmetries can be directly extracted from the
 symmetries generated by the pseudodifferential operators
 $A_{mn}$ of eq.~(\ref{eq2.36}) via Theorem~\ref{thm2}.
 Moreover, we will show that all the symmetries given in
 eq.~(\ref{eq2.40b}) that are local (no integrals), are
 point symmetries.
Bellow we shall call a symmetry trivial if it vanishes
 identically for any $ w(x,y) $

\begin{thm} \label{thm3}

a) Let $N$ be the number of an equation in the KP hierarchy.

b) $ m-n \le {N(1-n)} $;

c) all $ t_k \equiv  0 $ except $ t_0,t_1,t_2,t_N $~.

1. Then, for $n~ \ge ~0$ the $W_\infty \oplus H$ subalgebra
 of $P_\infty $ reduces to the positive part of the Lie
 point symmetries of the $KP_N$ equation.

2. For $n~\le~0$ the ``negative'' subalgebra $I_\infty ~
\subset ~ P_\infty $ reduces to the negative part of the
 Lie point symmetries of the $KP_N$ if we interpret ${\hat
x}^{-1}$ as in eq.~(\ref{i}).

\end{thm}

\begin{pf}
 (shortened). First let us rewrite the symmetries (\ref{eq2.40}) of the
 KP hierarchy using a different basis.  Instead of the
 operators $A_{mn} = \hat x^n \partial^m$ of eq.~
(\ref{eq2.36}), let us consider arbitrary functions
 $h_\alpha (x)$ (that can be expanded into a power series, or
 Laurent series).  We shall consider the
 operators
\begin{equation}
h_{\alpha, N}
	=~\lambda ^\alpha h_\alpha (\partial_E )~~,~~~
E(\lambda )~=~\lambda ^N~,\quad N \ne 0~;
\quad E(\lambda )= \ln \lambda ~,\quad
N=0~. \label{eq4.1}
\end{equation}
Sometimes we shall omit the label
$\alpha $ below.
We have the splitting of $h$ into the differential 
and the integral parts:
\begin{equation}
h~ =~ h_+~ +~ h_-~~,~~~ h_+~ =~ \sum_{n\ge 0} h_n \partial
_E^n~~ ,~~ h_-~=~ \sum_{n<0} h_n \partial_E^n~,
\label{eq4.2}
\end{equation}
where $h_n$ are constants. We rewrite the symmetries
 (\ref{eq2.40b}) in the form
\begin{equation}
V^N (\alpha, h)w(t_0,\vec t )
	= 2 \res_\lambda \lambda^\alpha
[ h(\partial _{E(\lambda )})
 \varphi(\lambda ,t_0,\vec t )] \varphi^* (\lambda
 ,t_0,\vec t )~, \label{eq4.3}
\end{equation}
where $V^N(\alpha, h)$ is a vector field acting on $w$ (a
 linear combination of the flows $V_{mn}$).

For $h(x) = (Nx)^n$, $\alpha = m + nN - n$, we recover the
symmetries (\ref{eq2.40b}) with $V^N(\alpha, h) = V_{mn}+ 
\sum_{k>0} c_k V_{m-k,n-k}$.

The condition (a) of {Theorem 3} is equivalent to the
 condition $\alpha\le N$ .

Let us rewrite eq.~(\ref{eq4.3}) in a more convenient form,
 using eq.~(\ref{eq2.30}) for the formal Baker-Akhiezer
 functions.  Our aim is to replace the power series in
 derivatives $\partial_\lambda$, implicit in eq.~
(\ref{eq4.3}), by power series in $\lambda$ itself.  The
 formula we are aiming at is
\begin{equation}
V^N(\alpha, h) w(t_0,\vec t )
	= 2 \res \lambda^\alpha f(\lambda ,t_0,\vec t )
\varphi(\lambda ,t_0,\vec t ) \varphi^* (\lambda ,t_0,\vec
 t )~, \label{eq4.4}
\end{equation}
where $f(\lambda ,t_0,\vec t
)$ is a function to be determined.  We shall use the
 following relation, valid for differential operators and
 (in view of (c)) also for integral ones:
\begin{equation}
h(\partial)e^\chi
	= e^\chi h(\partial + \chi') \cdot 1 = - e^\chi
\sum_{k=0}^\infty \frac
{h^{(k)}(\chi')}{k!}((\chi')^{-1}\partial)^k\cdot 1~ ,
\end{equation}
where $\chi' = \partial\chi/\partial x$~.

\noindent Comparing eq.~(\ref{eq4.3}) and eq.~(\ref{eq4.4})
 and using eq.~(\ref{eq2.30}) we have
\begin{mathletters} \label{eq4.6}
\begin{equation}
f(\lambda)
	= \varphi^{-1} (\lambda) h(\partial _E) e^\Phi~, 
\label{eq4.6a}
\end{equation}
\begin{equation}
\Phi
	= \sum_{k=1}^\infty \lambda^k t_k + 
\ln \biggl( 1 +
 \sum_{n=1} \lambda^{-n} K_n(t) \biggr)
+ t_0\ln \lambda~. \label{eq4.6b}
\end{equation}
\end{mathletters}\ignorespaces
The KP hierarchy is written in terms of the function $w =
 -2K_1$.  All higher coefficients $K_n$ in eq.~
(\ref{eq4.6b}) are expressed nonlocally in terms of $K_1$
 (see eq.~(\ref{eq2.15}), \dots ,(\ref{eq2.18})).  Since we
 will be using eq.~(\ref{eq4.4}) to extract {\em local}
 symmetries, we shall drop all negative powers
$\lambda^{-n}$ in eq.~(\ref{eq4.6b}), except for
$\lambda^{-1}$. We obtain
\begin{equation}
f(\lambda)
	= h \biggl\{ \frac{1}{N\lambda^{N-1}}
\frac{\partial}{\partial\lambda} + \sum_{k=1}^\infty
\frac{k}{N} \lambda^{k-N} t_k + \frac{1}{N}\lambda ^{-N}t_0
 + \frac{w}{2N\lambda^{N+1}} + O_1 \biggl(
\frac{1}{\lambda^{N+2}} \biggr) \biggr\} \cdot 1~.
\label{eq4.7}
\end{equation}
Another ingredient in eq.~(\ref{eq4.4}) is the expansion
(\ref{eq2.34}) which we rewrite as
\begin{equation}
2 \varphi \varphi^* =
	2 + \frac{w_x}{\lambda^2} + \frac{w_y}{\lambda^3} +
 \frac{w_t}{\lambda^4} + \dots +
\frac{w_{t_N}}{\lambda^{N+1}} + O_2 \biggl(
 \frac{1}{\lambda^{N+2}} \biggr)~. \label{eq4.8}
\end{equation}
 Both $O_1$ and $O_2$ contain nonlocal terms as coefficients
 before each degree of $\lambda^{-N-1-m}, m>0$. By
 comparing linear nonlocal terms for each degree of
 $\lambda$ in $O_1$ with similar terms in $O_2$ one can
 verify after some computation that they are different and
 therefore never cancel each other.

 Before expanding the function $h\{ \quad \}$ into a Taylor
 series we note that nonlocal terms (integrals of $w$) will
 be avoided iff the terms $O(\lambda^{-N-2})$ do not
 participate in the calculation of the residue in eq.~
(\ref{eq4.4}).  This imposes the necessary restriction
\begin{equation}
\alpha \le N~. \label{eq4.9}
\end{equation}
For the same reason, we must ``freeze'' all times in eq.~
(\ref{eq4.7}) except $t_0,t_1 = x$, $t_2 = y$ and $t_N$, where
 $N$ is the number of the chosen equation in the KP
 hierarchy that we are considering ($N=3$ for the KP
 itself, $N=4$ for $KP_4$ eq.~(\ref{eq2.26}) and so on).
Thus, we set
\begin{equation}
t_k = 0, \quad k \ge N+1, \quad 3 \le k \le N-1
\label{eq4.10}
\end{equation}
in eq.~(\ref{eq4.7}).

Finally, we expand $f(\lambda)$ in a Taylor series about
 $t_N$, keep only local terms and obtain for $N>2$
$$
V^N(\alpha ,h_\alpha )w
	= \res \lambda^\alpha \biggl\{ h_\alpha
(t_N) + \frac{1}{N}
 h_\alpha '(t_N) \biggl[ 2\lambda^{2-N} y + \lambda^{1-N}x +
\lambda^{-N}t_0 +  \frac{1}{2}\lambda^{-1-N}w \biggr]
$$
$$	+ \frac{1}{2!N^2} h_\alpha {''}(t_N) \biggl[
 4\lambda^{4-2N} y^2 + 4\lambda^{3-2N} xy + \lambda^{2-2N}
 (x^2 + 2y(2 + 2t_0 - N)) \biggr]
	+ \frac{1}{3!N^3} h_\alpha {'''} (t_N) \biggl[
 8y^3\lambda^{6-3N} + 12xy^2\lambda^{5-3N} \biggr]
$$
\begin{equation}
	+ \frac{1}{4!N^4} h_\alpha {''''}(t_N) 
16y^4\lambda^{8-4N}
 + O_1(\lambda^{-N-2}) \biggr\} \biggl\{ 2 +
 \frac{w_x}{\lambda^2} + \frac{w_y}{\lambda^3} + \dots +
 \frac{w_{t_N}}{\lambda^{N+1}} + O_2 (\lambda^{-N-2})
 \biggr\}~. \label{eq4.11}
\end{equation}

\end{pf}

Remark. One can continue~(\ref{eq4.11}) for $N=0,1,2$~.

\begin{thm} \label{thm4}
(Under the conditions of {Theorem 3}) The nonvanishing Lie
 point symmetries~(\ref{eq4.3}) are: (a) Virasoro
 symmetries for $\alpha = N$ (b) \ Kac-Moody symmetries
 $\alpha =2,1,-1$ (and also $\alpha=0$ for $N=3$). This
 algebra is a nilpotent one (c) The algebra of point
symmetries algebra is a semidirect sum of these two
 subalgebras.

\end{thm}

\begin{pf}
From~(\ref{eq4.11}) it follows that for $N>\alpha>2$
 symmetries vanish.
The fact that for $\alpha =N$ we get the Virasoro algebra
 (without central charge) follows from {Theorem 1}. This 
together with the
vanishing of symmetries for $N>\alpha>2$, completes the proof.
\end{pf}
Note that the values $N,2,1,0,-1$ are also the
``dimensions'' of $t_N,y,x,t_0,w$.
For five functions $h_N,h_2,h_1,h_0,h_{-1}$
we shall use the notations $f,g,h,k,\ell$ respectively.

Let us write down an analog of the Zakharov-Shabat
 representation for the point symmetry equations. Let us
 denote the group time corresponding to the flow
$V^N(\alpha ,h)$ as $z=z^N(\alpha ,h)$. From the
considerations in Section II one can obtain the symmetry
 equation as
\begin{equation}\label{ZSh}
[\partial _y - \partial ^2 - w_x~,~\partial _z + (L^\alpha
 h({\frac 1N}ML^{1-N}))_- ] = 0~,\quad N>0~,
\end{equation}
where for $L,M$ see~(\ref{LM}), for $M^{-1}$ see~(\ref{i}). The
 associated linear problem is the following one:
\begin{equation}\label{linearproblem}
(\partial _z - (L^\alpha h({\frac 1N}ML^{1-N}))_+)\varphi
 (\lambda ,t_0,{\vec t}) = \lambda ^\alpha h(\partial
_{\lambda ^N})\varphi (\lambda ,t_0,{\vec t})~,
\end{equation}
For $N=0$ we
take $h(ML),h(\lambda \partial _\lambda )$.

From Ref.~\cite{ref9} (see section $F$ in 
the present paper) and from a direct 
calculation it follows

\begin{thm} \label{thm5}
The point symmetries results from the 
following action on the $ \tau $-function:
\begin{equation}
V^N(\alpha,h)\tau(t_0,\vec t)=res_{\lambda}[\lambda
 ^{\alpha}[h(\partial_{\lambda^N})X(\lambda,
t_0,\vec t)]X^*(\lambda,t_0,\vec t)]\tau (t_0,\vec t)~,
\label{vertexpoint}
\end{equation}
$\alpha = N,2,1,0,-1$ ,
which is considered at the point $ \vec t =
 t_1,t_2,0,0,..,0,t_N,0,0,...~ $. For the 
integrtal part $h_-(\partial_{\lambda^N})$
the convention~(\ref{agreementi}) is used.

\end{thm}

Remark. In Ref~\cite{ref9} only the ``$\widehat 
{W}_{1+\infty}$ algebra'' action on the
 tau-function was considered.

\subsection{The PDO Lie algebra 2-cocycle and 
the free fermion algebra 2-cocycle}

Let us review the nice result of~\cite{ref32}, where an
 explicit expression for independent 
 nontrivial cocycles of the Lie algebra of the PDOs on
 the circle was obtained. In this subsection $\partial $ is
 $\partial _\lambda $. Given $A(\lambda ,\partial )~,~
B(\lambda ,\partial )~,~\lambda \in S^1$ 
are PDOs on the circle, these cocycles are

\begin{equation}\label{2cocycle}
\omega _1(A,B)~=~\frac {1}{2\pi i}~\oint ~\res _\partial ~
 A~[\ln \partial ~,B]~d\lambda \qquad,\qquad
\omega _2(A,B)~=~\frac {1}{2\pi i}~\oint~\res _\partial ~ A
~[\ln \lambda ~,B]~d\lambda~.
\end{equation}
Now let us solve the following problem: how to embed the
Lie algebra of the PDOs with a central extension into the
 free fermion algebra $\gl (\infty )$.
Let us give a short review of some facts from~\cite{ref7}.
One introduces free fermions

\begin{equation} \label{fermions}
\psi (\lambda ,t_0) = \sum _{n\in Z} \lambda ^{n+t_0}\psi _n~~
 ,~~ \psi ^*(\lambda ,t_0) = \sum _{n\in Z} \lambda
^{-n-t_0-1}\psi ^*~,
\end{equation}
where the fermion operators $\psi _n, \psi _n^* $~and
 vacuum are defined as in~\cite{ref7}.
One introduces the space $V$ of quadratic operators:
\begin{equation}\label{gl}
Q_A = \sum_{n,m=-\infty }^{+\infty }:\psi _n A_{nm}\psi
 _m^*:~,
\end{equation}
where $::$ denotes normal ordering and $A_{nm}$ is a
 generalized Jacobian matrix, i.e. infinite matrix with
 finite number of nonzero diagonals. The corresponding Lie
 algebra with the following commutation 
relation
\begin{equation}\label{iapcocycle}
[Q_A,Q_B]~ =~ Q_{[A,B]}~ +~ \omega 
_{gl(\infty)}(A,B)\quad,\quad
\omega _{gl(\infty)} (A,B)~=~ Tr[A,B]_{--}~ -~ 
Tr[A_{--},B_{--}]~,
\end{equation}
was called the $\widehat {\gl} (\infty )$ algebra.
In~(\ref{iapcocycle}) the subscript $--$ means a projection
of an infinite matrix $A_{nm}~n,m\in Z$ to its part $A_{nm}
~,~n,m<0~$. As in~\cite{ref9} we construct the embedding of
 differential operators into $V$ via~(\ref{gl}) and:
\begin{equation}\label{vlozhenie}
\lambda ~,~\lambda ^{-1} \to \Lambda ~,~\Lambda ^{-1} \quad
 ,\quad \partial _\lambda \to \Gamma \quad ,\quad (\Lambda
 )_{ik}~=~\delta _{i,k-1}~,~\Lambda ^{-1}~=~\delta
_{i,k+1}\quad , \quad  (\Gamma )_{ik}~=~
(k + t_0)\delta _{i-1,k}~.
\end{equation}
In~\cite{ref9} $t_0 \equiv 0$. Now we need noninteger $t_0$.
 Then we can also embed integral operators via:
\begin{equation}\label{integralop}
\partial _\lambda ^{-1}~\to~\Gamma ^{-1}~,
\end{equation}
where $\Gamma ^{-1}$ is defined for $t_0$ noninteger. The
 calculations provide the following result:
\begin{thm}\label{Theorem6}
For PDOs $A(\lambda ,\partial _\lambda )~,~B(\lambda
 ,\partial _\lambda )$ and embedding~
(\ref{vlozhenie}),(\ref{integralop}),(\ref{gl}) we have
\begin{equation}
\omega _{gl(\infty)}(A(\Lambda ,\Gamma ),B(\Lambda ,\Gamma ))
~\sim~
\omega _1(A(\lambda ,\partial ),B(\lambda ,\partial ))~+~
\omega _2(A(\lambda ,\partial ),B(\lambda ,\partial ))~,
\end{equation}
where $\omega_{1,2}$ are given by~(\ref{2cocycle}).
 Equivalently
\begin{equation}\label{nice!}
\omega _{gl(\infty)}(A(\Lambda ,\Gamma ),B(\Lambda ,\Gamma ))
~\sim~\res
 _{\lambda =\infty } \res _{\partial _\lambda} ~A(\lambda ,
 \partial )~[\sign (e)~,B(\lambda ,\partial)]~,
\end{equation}
where $e=\lambda \partial _\lambda$ is an Euler operator
 and $\sign(e) $ is a primitive of $\delta(e)$.
\end{thm}

The Lie algebra of PDO on the circle with this cocycle 
we denote by $\widehat {\cal A}_0$.

The formula~(\ref{nice!}) is suitable for practical
 calculations since only the delta function and its
 derivatives result from $\sign$ due to the commutator in
 this formula. Let us note that $\sign(e)~=~\sign(\Lambda
\Gamma ~-~t_0)$ looks like the
 $r$-matrix for $\gl(\infty )$.

Let us construct generators ($\alpha =N,2,1,0,-1;~
 E=E(\lambda )=\lambda ^N, N 
\neq 0;~E(\lambda )=\ln \lambda ,~N=0$):

\begin{equation}\label{fvirasoro}
L_{\alpha ,k}~=~res_\lambda ~ \lambda ^{\alpha -N} [(E\partial
_E^{k+1}+(j-1)(k+1)\partial _E^k)~
\psi (\lambda ,t_0)~]~\psi ^*(\lambda ,t_0)~ .
\end{equation}
We immediately get from~(\ref{gl}),(\ref{iapcocycle}) and
 Theorem 6  that
for $\alpha ~=~N$~this gives the Virasoro generators:

\begin{equation}\label{tvirasoro}
 [L_n,L_m]~=~(n-m)L_{m+n}~+~\frac {c_j}{12}~(n^3-n)\delta
_{m+n}~,\quad c_j~=~6j^2-6j+1~.
\end{equation}
Let us note that $L_{-1}~=~H_N$~(see (\ref{taufermion})
 below) and produces a higher $KP_N$ flow and $L_0$ for
 $j=0$ is the same as in~\cite{ref30} for Baker 0-forms.
Now let us remember the Sato definition of the tau
 function via free fermions:
\begin{equation}\label{taufermion}
\tau (\vec t)~=~<0\vert  ~\exp(\sum _{k>0}t_kH_k )g~\vert 0>
~,\quad H_k = Q_{\Lambda ^k}
= \sum _{n=-\infty }^{+\infty }\psi _n \psi
_{n+k}~,
\end{equation}
where $g$ is some element of Lie group $\widehat {GL}
 (\infty )$ of the Lie algebra $\widehat {\gl} (\infty )$.
 Let us introduce the following $\widehat {\gl} (\infty
 )$ flow on $\widehat {GL} (\infty )$ and hence the flow on
 the tau function:
\begin{equation}\label{tauflows}
\widehat {V}_{mn}g~=~Q_{\Lambda ^m\Gamma ^n}g \quad ,
\quad \widehat {V}_{mn}\tau~
=~<0\vert \exp(\sum _{k>0} t_kH_k) Q_{\Lambda ^m\Gamma
 ^n}g(t_{mn})\vert 0> ~ ,\quad n,m \in \Bbb Z~.
\end{equation}
The Lie algebra of these flows is the Lie algebra of PDOs with
 the central extension given by Theorem 6: $\widehat {V}_{mn}
\in \widehat {\cal A}_0$.
These flows on tau functions produce flows on formal
 Baker-Akhiezer functions $ \varphi ~,~ \varphi ^*~$. Let
 us remember
 the definition of these functions given in the papers of
 the Kyoto school~\cite{ref7}.
 Then the formal Zakharov-Shabat dressings which correspond
 to Baker-Akhiezer functions are uniquely reconstructed,
 $\tau \to K$,
 see Section II D.
\begin{thm}\label{Theorem7}
 Each flow $\widehat {V}_{mn}\tau $~(\ref{tauflows}) induces the 
 $P_\infty $ flow~$V_{mn}K$
~(\ref{eq2.37}),(\ref{eq2.39}) with the convention~
(\ref{ii}) on the space of formal
 Zakharov-Shabat dressings $\cal K $.
\end{thm}

 For $n \ge 0$ the flows~(\ref{tauflows}) were
 introduced before in~
\cite{ref9}, where the notation $\partial '_{mn} $ was 
used rather then $\widehat {V}_{mn}$. 
The algebra of these flows is known as the
 $\widehat {W} _{1+\infty }$ algebra, and a nice formula
 for its central extension is given by 
$\omega _{gl(\infty)} ~=~ \omega
 _1$ in~\cite{ref32}. We need $\omega
 _2$ for $I_{\infty }$ symmetries.

\subsection{The Virasoro subalgebra of point symmetries}

It is only the  part $n \le -1$ of Virasoro flows~
(\ref{fvirasoro},\ref{tvirasoro}) which reproduces the
point symmetries.
The other half of the Virasoro flows produces highly
 nonlocal expressions.
We can show that the difference between negative half of
 the point symmetries and the
 flows resulting from this Virasoro algebra action on
 fermionic tau-function
is a Liouville flow~\cite{ref33,ref38}. They describe a
 special case of Darboux transformations.

Let us obtain the Virasoro point symmetry algebra for each
 chosen equation in the KP hierarchy.
 We have in mind the equations written
 for the function $w(t_0,x,y,t,t_4, \dots) =
 -2K_1(t_0,x,y,t,t_4, \dots)$ in which all variables except
 $t_0, x, y$ and $t_N$ are ``frozen'': $\vec \epsilon = 0$.
  To obtain the Virasoro
 symmetries, we set $\alpha = N$ in eq.~(\ref{eq4.11}) and
 calculate the residue. The function $f $ below is $h_N $ of
 section A.  Let us consider each value of $N $ separately,
 starting formally from the cases $ N=0,1,2 $:
\begin{enumerate}
\item[0.] $N = \alpha = 0$~,\quad $\vec t =\vec \epsilon $.
\begin{equation}
T(f)=f'(t_0)\partial _w + O(\vec t)~.
\end{equation}

\item[1.] $N = \alpha = 1$~,\quad
$\vec t = x,\epsilon
 _2,\epsilon _3,... ~$..
\begin{equation}
T(f) = f(x)\partial - (f'(x)w + f''(x)(t_0^2-t_0))\partial
_w~+~O(\vec \epsilon )~.
\end{equation}
This formula can be interpreted as the known
 transformation of
a ``current'' $w(x)$ under the conformal transformation of
 space variable $x$. Then the term $f''(t_0^2-t_0)$ appears
 due to a central extension:
\begin{equation}\label{xVirasoro}
w_z = \{ w(x),{\cal T }_f \}~,\quad {\cal T }_f = \int
 f(x')(\frac 12 w^2(x') - (t_0^2-t_0)w'(x'))dx'~,\quad \{
 w(x),w(x') \} = \delta '(x-x')~.
\end{equation}

\item[2.] $N = \alpha = 2$~,\quad
$\vec t = x,y,\epsilon
_3,\epsilon _4,... ~$.
\begin{equation}\label{y}
T(f) = f(y)\partial _y +  \frac {f'(y)}{2}x\partial -
 (\frac {f'(y)w}{2} - \frac {x(1-2t_0)f''(y)}{4} + \frac
{x^3f'''(y)}{24})\partial _w
 + O(\vec \epsilon )~.
\end{equation}

\item[3.] $N = \alpha = 3$~,\quad $\vec
t = x,y,t,\epsilon
 _4,\epsilon _5,... ~$.

We obtain $T(f)~+ O(\vec \epsilon )$,
 where $T(f)$ as in eq.~
(\ref{KP3point1}),(\ref{KP3point2})

\item[4.] $N = \alpha = 4$~,\quad $\vec t = x,y,\epsilon
 _3,t_4,\epsilon _5,... ~$.

We obtain $T(f)~+ O(\vec \epsilon )$,
where $T(f)$ as in eq.~(\ref{KP4T}), $f
= f(t_4)$.

\item[5.] $N = \alpha = 5$,\quad $\vec t =
 x,y,\epsilon
 _3,\epsilon _4,t_5,\epsilon _6,... ~$.
\begin{equation}
T(f)
	= f \partial_{t_5} + \frac{1}{5} f'(x\partial_x +
2y\partial_y) - \frac{1}{25} (5wf' + 4y^2f{''}) \partial_w +
O(\vec \epsilon )~
, \quad f = f(t_5)~. \label{eq4.13}
\end{equation}

\item[6.] $N = \alpha \ge 6$,\quad $\vec t =
x,y,\epsilon
_3,...,\epsilon
_{N-1},t_N,\epsilon _{N+1},...
 ~$.
\begin{equation}
T(f)
	= f \partial_{t_N} + \frac{1}{N} f'(x\partial_x +
2y\partial_y) - \frac{1}{N} wf'\partial_w~, \quad f =
 f(t_N)~+
 O(\vec \epsilon )~. \label{eq4.14}
\end{equation}

\end{enumerate}

The commutation relations are the same in all cases, namely
 those of eq.~(\ref{eq3.4}). $O(\vec 0)=0$.

We note that the generators which correspond to the
 reparametrization of $y$~(\ref{y}) and $t_3$ variables
 create a rational KP potential~\cite{ref35}.

\subsection{The Kac-Moody subalgebras of point symmetries}
 The functions $g,h,k,\ell$ below are
 $h_\alpha $ from~(\ref{eq4.3}) and correspond to 
$\alpha =2,1,0,-1$
 respectively.

We obtain the following results.

$N = 0$, $~\alpha = -1$ yield the vector field

\begin{equation}\label{N0KacMoody}
U(\ell) = -2\ell (t_0)\partial_w~+ O(\vec t)~.
\end{equation}

$N = 1$, $~\alpha = 0,-1$ yield the vector fields

\begin{equation}\label{N1KacMoody}
W(k) = -2t_0  k'(x)\partial_w \quad, \quad
U(\ell ) = -2\ell (x)\partial_w~.
\end{equation}

$N = 2$, $~\alpha = 1,0,-1$ yield the vector fields

\begin{equation}\label{N2KacMoody}
X(h) = h(y)\partial_x - (t_0 h'(y) +\frac14
 h''(y)x^2)\partial_w~,~
W(k) = -k'(y)x\partial_w~,~
U(\ell) = -2\ell(y)\partial_w~.
\end{equation}

$N = 3$, $~\alpha = 2,1,0$ and $-1$ yield the vector fields  
$Y,X,W,U$ of~
(\ref{eq3.2}) respectively.

$N=4$, $\alpha = 2,1$ and $-1$ yield the vector fields
 $Y,X,U$ of (\ref{KP4X}),
 respectively.

$N = 5$, $~\alpha = 2,1$ and $-1$ lead to
\begin{equation}\label{eq4.15}
Y(g)
	= g(t_5)\partial_y - \frac{4}{5} g'y\partial_w \quad, 
\quad X(h)
	= h(t_5)\partial_x \quad ,\quad 
U(\ell)
	= -2\ell(t_5)\partial_w~,
\end{equation}
respectively.

$N \ge 6$, $~\alpha = 2, 1$ and $-1$ lead respectively to
\begin{equation}\label{eq4.16}
Y(g)
	= g(t_N)\partial_y \quad, \quad
X(h)
	= h(t_N)\partial_x \quad,\quad
U(\ell)
	= -2\ell(t_N)\partial_w~. 
\end{equation}

\subsection{The symmetry groups}
The advantage of point symmetries is that they can be
 easily and explicitily be integrated to give the group
 transformations. Let us construct the corresponding
 transformations by integrating the one parameter
 subalgebras of the vector fields (\ref{point}), obtained
 above (with $\vec \epsilon = 0$). For the sake of simplicity 
 we shall use the same
 notation $z $  for different group times. The functions
 $f,g,h,k,\ell$ below are $h_\alpha $
from~(\ref{eq4.3}) for $\alpha =N,2,1,0,-1$ respectively, see
 sections $B$,$C$.

\begin{mathletters}\label{eq4.17}
\begin{equation}
\frac{d \tilde x}{dz} 
	= v_1(\tilde x, \tilde y, \tilde t_N, \tilde w)
\quad, \quad \frac{d \tilde y}{dz} 
	= v_2(\tilde x, \tilde y, \tilde t_N, \tilde w)
\quad,\quad 
\frac{d {\tilde t}_N}{dz} 
	= v_N(\tilde x, \tilde y, \tilde t_N, \tilde w)
\quad, \quad \frac{d \tilde w}{dz} 
	= p(\tilde x, \tilde y, \tilde t_N, \tilde w)~,
 \label{eq4.17a}
\end{equation}

\begin{equation}  
\tilde x \big|_{z=0} = x\quad, \quad \tilde y \big|_{z=0} =
 y\quad,
\quad \tilde t_N \big|_{z=0} = t_N\quad, \quad \tilde w 
\big|_{z=0} = w \quad,\quad \frac {\partial \tilde t
 _N}{\partial t _N}~=~\frac {f(\tilde t _N)}{f(t _N)}~.
 \label{eq4.17b}
\end{equation}

\end{mathletters}\ignorespaces

\begin{description}
\item[$N \ge 6$].   Virasoro (eq.~(\ref{eq4.14})):

\begin{mathletters}\label{eq4.18}
\begin{equation}
\tilde t_N
	= \phi^{-1}\bigr(z +\phi (t_N)\bigr), \quad
\tilde x
	= x \biggl( \frac{\partial \tilde t_N}{\partial
t_N}\biggr)^{1/N} , \quad
\tilde y
	= y \biggl( \frac{\partial \tilde t_N}{\partial
 t_N} \biggr)^{2/N}, \quad
\tilde w
	= w \biggl( \frac{\partial \tilde t_N}{\partial
 t_N} \biggr)^{-1/N},\quad \phi(t)=
\int^t\frac{dt'}{f(t')}.
 \label{eq4.18a}
\end{equation}

Thus we see that under the change of~ $t_N $~ the variables
 $x,~y,~w~$ transform as $\frac {1}{N},~\frac {2}{N}$,~and
 $-\frac{1}{N} $ tenzors respectively.
The Kac-Moody algebra (eq.(\ref{eq4.16})) integrates to:
\begin{equation}
\tilde t _N = t_N \quad, \quad \tilde x = x + z h(t_N)
\quad, \quad \tilde y = y + z g(t_N) \quad, \quad \tilde w
 = w + 2z\ell(t_N)~. \label{eq4.18b}
\end{equation}
\end{mathletters}\ignorespaces

\item[$N = 5$].  Virasoro (eq.~(\ref{eq4.13})):
\begin{mathletters}
\begin{equation}
\tilde t_5 = \phi^{-1}\bigl(z + \phi(t_5)\bigr)~, \quad 
\tilde x
= x \biggl( \frac{\partial \tilde t_5}{\partial t_5}
 \biggr)^{1/5}~, \quad \tilde y = y \biggl( \frac{\partial
 \tilde t_5}{\partial t_5} \biggr)^{2/5}~,\quad
 \tilde w = \biggl( \frac{\partial \tilde t_5}{\partial t
 _5} \biggr)^{-1/5} \biggl[ w - \frac{4}{25} y^2 \frac{\dot
 f(\tilde t _5) - \dot f(t_5)}{f(t_5)} \biggr]~.
 \label{eq4.19a}
\end{equation}

Kac-Moody  (eq.(\ref{eq4.15}))
\begin{equation}
\tilde t _5 = t_5~, \quad \tilde x = x + z h(t_5)~, \quad
 \tilde y = y + z g(t_5)~, \quad \tilde w = w + 
2\ell(t_5)z - \frac{4}{5}
g'(t_5) y z - \frac{2}{5} g(t_5) g'(t_5) z^2~.
\label{eq4.19b}
\end{equation}
\end{mathletters}\ignorespaces

\item[$N=4$] Virasoro (eq.~(\ref{KP4T})):
\begin{mathletters}
\begin{equation}
\tilde t_4 = \phi^{-1}\bigl(z + \phi(t_4) \bigr)~, \quad
	\tilde x = x \biggl( \frac{\partial \tilde
 t_4}{\partial t_4}\biggr)^{1/4}~, \quad \tilde y =
 y\biggl( \frac{\partial \tilde t_4}{\partial
 t_4}\biggr)^{1/2}~, \quad
\tilde w =\biggl( \frac{\partial \tilde t_4}{\partial
 t_4}\biggr)^{-1/4}
\biggl( w - \frac{1}{4} xy \frac{\dot f(\tilde t_4) -
 \dot f(t_4)}{f(t_4)} \biggr)~. \label{eq4.20a}
\end{equation}
\noindent Kac-Moody (eq.~(\ref{KP4X})):
$$
\tilde t _4 = t_4~, \quad \tilde x = x + z h(t_4)~, \quad
\tilde y = y + z g(t_4)~,
$$
\begin{equation}
\tilde w = w +\biggl(2\ell(t_4) - \frac{1}{2} x\dot g(t_4) -
 y \dot h(t_4) \biggr)z - \biggl( \dot g(t_4) h(t_4) +
 2g(t_4) \dot h(t_4) \biggr) \frac{z^2}{4}~. \label{eq4.20b}
\end{equation}
\end{mathletters}\ignorespaces
\end{description}
The formulas for the group transformations for $N=3$ (the
 KP equation itself) are somewhat more complicated.  They
 were given in Ref.~\cite{ref21,ref22} and we do not
 reproduce them here.
\begin{description}
\item[$N=2$.]Virasoro (eq.~(\ref{y}))
$$
\tilde y =\phi^{-1}\bigl(z+\phi(y)\bigr),\quad \tilde x =
x \biggl(\frac{\partial \tilde y}{\partial y}\biggr)^{1/2}
$$
\begin{eqnarray*}
\tilde w(\tilde x,\tilde y) &=& \biggl(\frac{\partial \tilde
  y}{\partial y}\biggr)^{-1/2}w(x,y) + 
\frac{\tilde x(1-2t_0)}{f(\tilde
  y)}[f'(\tilde y)-f'(y)]\\
& & +\frac{\tilde x^3}{24f(\tilde y)^2}\{f(y)f''(y)-f(\tilde
y)f''(\tilde y)\}+\frac 12
[f^{\prime 2}(\tilde y)-f^{\prime 2}(y)]
\end{eqnarray*}
Kac Moody (eq.~(\ref{N2KacMoody}))
$$
\tilde y=y, \quad \tilde x = x +zh(y)
$$
$$
\tilde w(\tilde x,\tilde y) = w(x,y) - t_0h'(y)z-\frac14
h''(y)[x^2z+xh(y)z +\frac 13 h^2(y)z^3]
$$
\end{description}
The Virasoro algebra induces a reparametrization
 of time $t_N$, that is compensated for by a
 redefinition of the other variables.

\subsection{Restriction to integrable equations in
 (1+1)-dimensions. Compatible pairs and flows}

First we should mark that a large number of finite
 dimensional systems and integrable $1+1 $-dimensional
 systems can be obtained from the $KP$ hierarchy or the
 vector fields considered in Chapter II by imposing
 constraints.

Let us impose the following constraint on the space of the
 formal Zakharov-Shabat dressings ${\cal K} $: given a PDO
 $A_i $ of order $p_i$ ,  we set that $(KA_iK^{-1})_- = L_{i~-}
 $~ is a Volterra integral operator with degenerate kernel
 of rank $r_i<\infty $, i.e.:

\begin{equation} \label{rank}
L_i~=~KA_iK^{-1}~,\quad K \in {\cal K}  ~: \quad    rank~
(KA_iK^{-1})_-~=~ r_i~<~\infty ~.
\end{equation}
\newtheorem{lemmarank}{Lemma}
\begin{lemmarank}
For $p_i>0$ this constraint restricts the space $\cal K$ to
 the subspace ~$ {\cal K}  [A_i]~\subset \cal K $ which is
 parametrized by the set of $p_i~+~2r_i~-~1$ arbitrary
 functions of $x$.
\end{lemmarank}

 To get a finite dimensional system (an ordinary
 differential equation),  let us consider the $pair$ of
 constraints~(\ref{rank}) $i=1,2$. The question is: does
 any $K \in \cal K$ solving both constraints exist? If it
 exists then we can consider the algebra of constraints
 spanned by $\{ A_i~,~i=1,2\} $. Example 1: both $A_i$ do
 not depend on $x$. When the solution of~(\ref{rank})
 exists it is the Burchnall-Chaundy commutative ring~
\cite{ref40} spanned by $\{ L_i~,~i=1,2\} $. Example 2: the
 string equations~\cite{ref41}\cite{ref46},
 see~
(\ref{virflows}),(\ref{hernia}) for$A_i=Q,P,\hat P$.
 Example 3: the bispectral problem~\cite{ref42}.

\begin{problem}\label{problem1}
To classify all pairs of PDOs $A_i$ with finite $r_i$ which
 can be simultaneously transformed by a formal
 Zakharov-Shabat dressing $K \in \cal K $ to
\begin{equation}
\{ A_1~,~A_2\} \quad \to ^K \quad \{L_1~,~L_2\} \qquad :
\quad  rank~ L_{i~-}~=~r_i~,\quad i=1,2~,
\end{equation}
and to describe the corresponding subspace of the formal
 Zakharov-Shabat dressings ${\cal K}[A_1,A_2] \subset \cal
 K$.
\end{problem}

Let us denote the subalgebra of PDOs spanned by $ A_i~,~
i=1,2$ by ${\cal A}[A_1,A_2]$.   Let us make a simple
 proposition:
\begin{proposition}
For $r_1,r_2=0$ the space ${\cal K}[A_1,A_2]$ is empty if
 ${\cal A}[A_1,A_2]$ contains purely integral operators.
\end{proposition}
Example. p-KdV equation is defined by
\begin{equation}\label{p-KdV}
(K\partial ^p K^{-1})_-~=~0~.
\end{equation}
It is impossible to impose the following Virasoro
 (``string-type'') constraint $(K{\hat x}\partial
 ^q K^{-1})_-=0$ for p-KdV equation if $p+q-1<0$.

~

To get 1+1 dimensional integrable systems like the KdV or
 NLS equations one needs 1) To impose one constraint~
(\ref{rank}) 2) To choose the vector field~(\ref{eq2.10}).
Again the question is whether the given flow~(\ref{eq2.10})
 is compatible with the given constraint~(\ref{rank}).

For $L_i~=~KA_iK^{-1}$~and the flow $V_j $~(\ref{eq2.10})
 we have:
\begin{equation} \label{Lax}
V_j L_i~=~[~L_i~,~(KA_jK^{-1})_-~]~=~ [~(KA_jK^{-1})_+~,~
L_i ~]~+~KF_{ij}K^{-1}~,\quad
F_{ij}~=~[A_i,A_j]~+~V_jA_i~.
\end{equation}

\begin{problem}\label{problem2}
To describe all flows~(\ref{eq2.10}) which preserve a given
 constraint~(\ref{rank}).
\end{problem}

If the curvature $F_{ij}=0$ then the flow preserves the
 constraint~(\ref{rank}). Then~(\ref{Lax}) is a Lax system
 of equations for $p_i-1+2r_i$ independent functions in the
 two dimensional space spanned by ~$~x~,~z_j~$.
Examples~\cite{ref33,ref38}. If we take $A_i=\partial ~,~
A_j=\partial ^2 $ we get the $r_i $-component nonlinear
Schroedinger equation. By taking $x$-independent PDOs $A_i
 $ of higher order we got systems which were called
 ``constrained KP hierarchies''. Certain pairs of such
 restrictions give rise to different types of restricted
 flows~\cite{ref37}, which are either autonomous or
 nonautonomous finite-dimensional dynamical systems.

~

Let us consider the problem of the compatibility of the
 KdV-type constraints with any $KP$ time dependent
 symmetry. This includes point symmetries as a particular
 case.

By the KdV-type constraint we mean the constraint:

\begin{equation} \label{constraint}
(KA_0K^{-1})_- = 0 ~,
\end{equation}
where $A_0$ is a PDO of order $p_0>0$ which does not depend
 on $x$.

As we know from the Lemma in Section II, if the order of $
 A_0=p_0< \infty $ then the constraint~(\ref{constraint})
 restricts the space of formal Zakharov-Shabat dressings $
 {\cal K} $ to the set of $ p-1 $ functions $ {\cal K} =
 \{ K_j , j=1,2,...,p-1 \} $ .

Then the condition of compatibility of this KdV-type
 constraint
with $P_\infty $ symmetry action produced by $V_i$ with
 time $z_i$, where $z_i$  is the corresponding time due to
 the Theorem 1 is

\begin{equation} \label{condition}
(KF_{0i}K^{-1})_- = 0~~~ ,~~~F_{0i} = [A_0,A_i] + \frac
 {\partial A_0}{\partial z_i}~=~[\nabla _i~,~A_0].
\end{equation}
and the same we is obtained for each polynomial of $ A_0$.
 These are additional constraints.  We continue this
 process and finally get some space of constraints that we
 denote by ${\cal A}(A_0;\nabla _i)$ (where we put $\nabla
_i = \partial _{z_i}-A_i$) resulting from original
 constraint and the given flow. Generally we shall denote
 by ${\cal A}(~\vec A~ ;~ \vec {\nabla }~ )$ the algebra of
 constraints which one gets by imposing a set of
 constraints~(\ref{constraint}) with $\vec A~=~A_0,A_1,...
$, and vector fields $\vec V~=V_i,V_k,..$ corresponding to
   $\vec \nabla ~=~\nabla _j,\nabla _k,...$ by taking all
 products and sums of constraints and by commuting them
 with all ``covariant derivatives''. As a result ${\cal A}(
~\vec A~;~\vec \nabla ~) $ is stable under multiplying by
 any constraint and under commuting with any $\nabla $. The
 corresponding subspace of $\cal K$
which transform each element of ${\cal A}(~\vec A~;~\vec
 \nabla ~) $ to a differential operator we denote by ${\cal
 K }(A_0;A_i-\partial _{z_i})$.

We obtain a simple proposition:
\begin{proposition}
~~(a) If~ ${\cal A}(A_0;A_i-\partial_{z_i}) \subseteq {\cal
 A}(A_0)$~ then the flow with respect to $z_i$ is
 compatible with the constraint.~
~(b)~If~ ${\cal A}(A_0;A_i-\partial _{z_i})$ contains a
 nonzero element which belongs to the purely integral
 operators, then ${\cal K }(A_0;A_i-\partial _{z_i})$ is
 empty.
\end{proposition}

For the case of general PDO $A_0$ and $\nabla $~we can
 formulate the following

\newtheorem{gipoteza}{Conjecture}
\begin{gipoteza}
If~ ${\cal A}(A_0;\nabla _ i)$ does not satisfy neither
 condition (a) nor (b) of the previous Theorem then we get
 a further restriction of our space ${\cal K } $ - it is
 finite dimensional.
\end{gipoteza}

Bellow in the case under consideration this Conjecture can
 be proved via Theorem 1 .

Example. $p-KdV$ constraint:

\begin{equation} \label{KdV}
(K\partial ^pK^{-1})_-=0
\end{equation}

We have $ {\cal K}=\{ K_j(x),~ j=1,2,...,p-1 \} $. Let us
 treat the $KP$ flow given by $A_i=\partial ^i$. It is easy
 to see that ${\cal A}(\partial ^p;\partial^i-\partial_{t_i})
={\cal A}(\partial^p)$, it means the known fact that all 
 $ KP $ flows preserve the
 $ p-KdV $ reduction. Another example is the compatibility
 of Virasoro flows from~\cite{ref9} with $ p-KdV $
 reduction:

\begin{equation}
{\cal A}(\partial ^p;\partial _{q1}-\hat x\partial ^q)=
{\cal A}(\partial ^{(pn + (q-1)m)})~,~n>0,m\ge 0 .
\end{equation}

Hence it is only for $q=1-p+kp, k>0$ that the Virasoro flow
 is compatible with $p-KdV$.
For $q<1-p~$  Theorem 6 implies that there are no solutions.
 For different values of $q$ we get the $conditional~
symmetry$ for the finite-dimensional system corresponding
 to the rational curve finite-gap solution. This is in the
 accordance with the Conjecture.

From the above it follows that
$$
 Only~ linearly~ x~-dependent~ A_j~ are~ compatible~ with~
 the~ KdV-type~ constraint~(\ref{constraint}) .~ These~ are
 ~A_j~=~{\hat x} (A_0)^k(A_0 ')^{-1}.
$$

The only possibility to get point symmetries is to treat
 flows corresponding to $x$-dependent $A_j$ or
 corresponding to simple shifts in $x$ and time variables.
 Thus we get simple shifts in $x$ and time, and scaling and
 Galilei transformations only. This explains why only finite
 dimensional algebras of point symmetries are obtained by
 direct methods~\cite{ref21},\cite{ref22}.

\section{Conclusions}
We have described $P_\infty $ symmetries of integrable
 systems. We found
the classical symmetries of Sophus Lie (point
symmetries) of all higher $KP$ equations as an example.
 We
proved that each time the symmetry algebra is a
semidirect sum of the Virasoro and current algebras.
Each Virasoro algebra corresponds to the reparametrization group
 of the corresponding  $KP$ higher time. The topic is
 embedded into the standard solitonic theory, like $L-A$
 pairs, tau-functions, $W_\infty $ symmetries.

Let us mention some related problems which were  not
 considered in this paper. We noted in the Introduction
that the notion of point symmetries depends on the
 choice of variables and representations for $KP$ flows.
In soliton theory different variables can be
connected in a highly nonlocal way. (Moreover, soliton
theory may be viewed as a science about parametrizations
and changes of variables, and in fact in this quality
has a lot of applications in both mathematics and
theoretical physics). The point transformations have
the advantage of being easily integrated to the group
transformations. Therefore for possible applications
like Ward identities it may appear to be suitable to
have a ``point representation'' for each
symmetry of the system under investigation.
 The problem is: given a $\widehat {gl}(\infty )$ or
$P_\infty $ symmetry flow, what are the variables, for
which  this flow corresponds to a point transformation?
Does each element of a $KP~-~\widehat {gl}(\infty )$ or
$KP~-~P_\infty $ algebra gives rise to a certain point
symmetry?

We can suggest some possible applications of these
symmetries.
First, the role of symmetries for the quantization
problem is known.
Let us note that the ``spectral parameter
reparametrization''~\cite{ref11,ref30} Virasoro
algebra applications appeared nondirectly
\cite{ref13,ref39,ref2,ref16}. This algebra was not
connected with reparametrizations of the space variable
$x$, which is of importance in conformal theories of
phase transitions in 2D. One can expect the appearance
of Virasoro algebra in the solvable models arising
from space variables reparametrizations. We are sure
that the ``time reparimetrization'' Virasoro will be of
use.

A different application is to use these
reparametrization symmetries for considering integrable
equations on Riemann surfaces, where one needs change
of $t$-variable to glue different maps.

The other application is the standard one: the
calculation of special solutions.

An interesting and very complicated problem is posed in
the last part of the paper: to classify all compatible
pairs $A_i,A_j$ and corresponding subspaces of the
formal Zakharov-Shabat dressings ${\cal K}[A_i,A_j]$.
Along with~\cite{ref6} we can expect that all ordinary
equations we get in such a way have the Painleve
property. Is the reverse statement true? What sort of
special functions can be obtained
in such a way? What sort of representation theory is
connected with these special
solutions? About applications of~\cite{ref18} see
\cite{ref17}. As for $gl(\infty )$ part of symmetries
we know each time the correspondence between free
fermion algebra and differential equations. What to
do with the $I_\infty $ part which has no free fermion
interpretation for the convention~(\ref{i})?

The way of getting point symmetries described in this
paper is available not only for the $KP$ hierarchy
but for the multicomponent $KP$ hierarchy and therefore
 almost for all known integrable models, like the 3D 3 wave
 resonant system hierarchy, Toda lattice hierarchy,
 Davey-Stewartson one e.t.c.
It is also known the continuous generalization of
 the multicomponent KP equation, where t-dependent
  symmetries
 were found via recursion operator method Ref~\cite{ref45}.
 The results of this article
 confirm a conjecture made earlier that the Lie point
 symmetries of integrable PDEs in 3-dimensions are
 infinite-dimensional and have ``a characteristic Kac-Moody
 structure''~\cite{ref25}. Since Lie point symmetries can
 be found using simple algorithms and computer packages~
\cite{ref20}, they provide a tool for investigating any
 given equation, and excluding it from the list of
 integrable equations originating from the $KP$ in case
there is no infinite-dimensional symmetry group with the
 prescribed structure.

\section*{Acknowledgements}   The research of one of the
 authors (P.W.) was partially supported by research grants
 from NSERC of Canada and FCAR du Qu\'ebec.  A. Yu. O.
 thanks the RFFI (grant 96-02-19085), INTAS and
NSERC of Canada for support.


\begin{references}
\bibitem{ref1} V. V. Kadomtsev and V. I. Petviashvili,
Soviet Phys. Dokl. {\bf 15}, 539 (1970).
\bibitem{ref2} A.Yu.Morozov, ``Integrability and Matrix
 Models'', Uspehi Fizicheskih Nauk, vol 164 (1994) pp.3-62
\bibitem{ref3} V. Dryuma, Soviet Phys. JETP {\bf 19}, 387
 (1974).
\bibitem{ref4} V. E. Zakharov and A. B. Shabat, J. Funct.
 Anal. Appl. {\bf 8}, 226 (1974), {\bf 13}, 166 (1979).
\bibitem{ref5} V.E.Zakharov, S.V.Manakov, S.P.Novikov,
 L.P.Pitaevsky {\em The Theory of Solitons. 
The Inverse
 Scattering Method}.
\bibitem{ref6} M. J. Ablowitz and P. Clarkson, Solitons,
 Nonlinear Evolution Equations and Inverse Scattering,
 Cambridge University Press, Cambridge, 1991.
\bibitem{ref7} M. Jimbo and T. Miwa, Publ. RIMS Kyoto Univ.
 {\bf 19}, 943 (1983).
\bibitem{ref8} L. A. Dickey, {\em Soliton Equations and
 Hamiltonian System}, World Scientific, Singapore, 1991.
\bibitem{ref9} A. Yu Orlov, ``Vertex Operators, $\partial $
 bar Problem, Symmetries, Hamiltonian and Lagrangian
 Formalism of (2+1) Dimensional Integrable Systems'', in
  {\em Plazma Theory and Nonlinear and Turbulent Processes in
 Physics, Proc. III Kiev. Intern. Workshop (1987)}, vol.
 I, pp. 116--134, World Scientific Pub, Singapore, 1988, (editors:
 V.G.Bar'yakhtar,...,V.E.Zakharov).
\bibitem{ref10} A. Yu. Orlov and E. I. Schulman, Additional
 Symmetries of Two Dimensional Integrable Systems (in
 Russian), Preprint 277, Novosibirsk, 1985.
\bibitem{ref11} A. Yu. Orlov and E. I. Schulman,
 Teoreticheskaya i Matematicheskaya Fizika, {\bf 64}, 323
(1985); ``Additional Symmetries of 1+1 Dimensional
 Integrable Equations and Conformal Algebra
 Representation'' Preprint 217, Novosibirsk, (1984).
\bibitem{ref12} A. Yu Orlov and E. I. Schulman, Lett. Math.
 Phys. {\bf 12} (1986) p.171.
\bibitem{ref13} A. Gerasimov, A. Marshakov, A. Mironov,
A. Morozov, and A. Orlov, Nuclear Phys. {\bf B 257},  565 (1991).
\bibitem{ref14} A. Gerasimov, Yu. Makeenko, A. Marshakov,
 A. Mironov, A. Morozov, and A. Orlov, Mod. Phys. Lett.
 {\bf A6}, 3079 (1991).
\bibitem{ref15} S.Kharchev, A.Marshakov, A.Mironov,
A.Orlov, A.Zabrodin, Nuclear Physics {\bf B 366}, 
pp.569-601 (1991)
\bibitem{ref16} K.Takasaki, Communications in Mathematical
 Physycs, {\bf 181} pp.131-156 (1996); 
hep-th/9506089, 35 pp
\bibitem{ref17} B.Bakalov, E.Horozov, M.Yakimov, ``Highest
 weight modules over $ W_{1+\infty} $ algebra and the
 bispectral problem'', Sofia University preprint, 1996.
\bibitem{ref18} V.Kac, A.Radul, Communications in
 Mathematical Physics, 157 (1993), pp. 429-457.
\bibitem{ref19} P. J. Olver, {\em Applications of Lie
 Groups to Differential Equations} (Springer, Berlin, 1986).
\bibitem{ref20} B. Champagne, W. Hereman and P. Winternitz,
 Comput. Phys. Comm., {\bf 66}, 319 (1991).
\bibitem{ref21} D. David, N. Kamran, D. Levi, and P.
 Winternitz, Phys. Rev. Lett. {\bf 55}, 2111 (1985), J.
 Math. Phys. {\bf 27}, 1225 (1986).
\bibitem{ref22} D. David, D. Levi and P. Winternitz, Phys.
 Lett. {\bf A118}, 390 (1986).
\bibitem{ref23} B. Champagne and P. Winternitz, J. Math.
 Phys. {\bf 29}, 1 (1988).
\bibitem{ref24} L. Martina and P. Winternitz, Ann. Physics
 (N.Y.) {\bf 196}, 231 (1989).
\bibitem{ref25} P. Winternitz, in {\em Symmetries and
 Nonlinear Phenomena}, World Scientific. Singapore,
358--375 (editors: D. Levi and P. Winternitz).
\bibitem{ref26} B. Dorizzi, B. Grammaticos, A. Ramani,
 and P. Winternitz, J. Math. Phys. {\bf 27}, 2848 (1986).
\bibitem{ref27} J. Rubin and P. Winternitz, J. Math. Phys.
 {\bf 31}, 2085 (1990).
\bibitem{ref28} D. David, D. Levi and P. Winternitz, Phys.
 Lett. {\bf A129} 161 (1988).
\bibitem{ref29} D. Levi and P. Winternitz, J. Math. Phys.
 {\bf 34}, 3713 (1993).
\bibitem{ref30} P.G.Grinevich, A.Yu.Orlov, ``Virasoro
 Action on Riemann Surfaces, Grassmannians,  $ det {\bar
 \partial }_j $ and Segal-Wilson tau-function'', in
 {\em Problems of Modern Quantum Field Theory} pp.
 86-106, Springer, 1989
 (editors: A.A.Belavin, A.U.Klimuk, A.B.Zamolodchikov).
\bibitem{ref31} P. G. Grinevich and A. Yu. Orlov,
 Funktsional. Anal. i Prilozhen. {\bf 24}, 72 (1990).
\bibitem{ref32} O.S.Kravchenko, B.A.Khesin, Funkt. Analiz i
 Ego Prilozheniya,
{\bf 25}, No 2  (1991)
\bibitem{ref33} A. Yu. Orlov, 
``Volterra Operators for Zero Curvature 
Representation. Universality of KP'', 
in  {\em Plazma Theory and  Nonlinear and Turbulent 
 Processes in Physics}, pp. 126-131, Springer Series in 
Nonlinear Dynamics, 1991 (editors: A.Fokas, 
D.Kaup, A.Newell, V.E.Zakharov).
\bibitem{ref34} G.Segal, G.Wilson, Publ. Math. IHES 
{\bf 61}, 5
 (1985).
\bibitem{ref35} A.P.Veselov, Communications of the Moscow
Mathematical Society, Russian Math Survay {\bf 35:1}
 (1980) pp. 239-240.
\bibitem{ref36} K. Ueno and K. Takasaki, Adv. Stud. Pure
Math. {\bf 4}, 1 (1984).
\bibitem{ref37} A.Yu.Orlov, S.Rauch-Wojciechowski, Physica
 {\bf D 69} (1993).
\bibitem{ref38} A.Yu.Orlov, ``Symmetries for Unifying
 Different Soliton Systems into a Single Integrable
 Hierarchy'', preprint IINS/Oce-04/03, March 1991; the
report at the NEEDS workshop, Gallipolli 1991.
\bibitem{ref39} P. van Moerbeke, {\em Integrable
 Foundations of String Theory}, Lectures at CIMPA Summer
 School, Nice, 1991.
\bibitem{ref40} J.L.Burchnall, T.W.Chaundy, Proc. Royal
 Soc. A, {\bf 118} (1928) pp. 557-583.
\bibitem{ref41} M.Adler, P. van Moerbeke, ``
Compatible Poisson structures and the Virasoro 
algebra'' (1995)
\bibitem{ref42} J.J.Duistermaat, F.A.Grundbaum, Comm. Math.
 Phys. {\bf 103} (1986) pp.177-240.
\bibitem{ref43} L.A.Dickey, Modern Physics 
Letters A, {\bf 8}, No 13 (1993) pp.1259-1272;
 {\bf 8}, No 14 (1993) pp.1357-1377
\bibitem{ref44} A.Yu.Orlov, P.Winternitz, ''Algebra of
 pseudodifferential operators and symmetries of equations
 in the Kadomtsev-Petviashvili hierarchy''
\bibitem{ref45} A.Fokas and P. Santini, Comm. Math.
 Phys. {\bf 116} (1989) pp.449-474.
\bibitem{ref46} T.Shiota, Comm. Math.
 Phys. (1995)

\end{references}
\end{document}